\documentclass[aps,prd,onecolumn,groupedaddress,showpacs,nofootinbib,amssymb]{revtex4}
\usepackage[dvips]{graphicx}
\usepackage{amssymb}
\usepackage{amsmath}
\usepackage{graphicx,,color}
\usepackage{amsfonts}
\usepackage{bm}
\usepackage{cancel}
\usepackage{comment}

\newcommand\be{\begin{equation}}
\newcommand\ee{\end{equation}}

\allowdisplaybreaks[4]

\begin{document}

\tolerance=5000

\title{Unification of a Bounce with a Viable Dark Energy Era in Gauss-Bonnet Gravity}
\author{S.D.~Odintsov,$^{1,2}$\,\thanks{odintsov@ice.cat}
V.K.~Oikonomou,$^{3,4}$\,\thanks{v.k.oikonomou1979@gmail.com}F.P.
Fronimos,$^{3}$\,\thanks{fotisfronimos@gmail.com}K.V.
Fasoulakos,$^{3}$\,\thanks{kvfasoulakos@gmail.com}}
\affiliation{$^{1)}$ ICREA, Passeig Luis Companys, 23, 08010 Barcelona, Spain\\
$^{2)}$ Institute of Space Sciences (IEEC-CSIC) C. Can Magrans
s/n,
08193 Barcelona, Spain\\
$^{3)}$ Department of Physics, Aristotle University of
Thessaloniki, Thessaloniki 54124,
Greece\\
$^{4)}$ Laboratory for Theoretical Cosmology, Tomsk State
University of Control Systems and Radioelectronics, 634050 Tomsk,
Russia (TUSUR)}

\tolerance=5000

\begin{abstract}
In this work we shall demonstrate that it is possible to describe
in a unified way a primordial bounce with the dark energy era, in
the context of Gauss-Bonnet modified gravity. Particularly, the
early time bounce has a nearly scale invariant power spectrum of
primordial scalar curvature perturbations, while the dark energy
era is a viable one, meaning that it mimics the
$\Lambda$-Cold-Dark-Matter model and also is compatible with the
Planck 2018 data on cosmological parameters. In addition, our
analysis indicates that the dark energy era is free from dark
energy oscillations, which occur in the context of $f(R)$ gravity.
We further addressed the later issue by examining $f(R)$
extensions of Gauss-Bonnet models, and we showed that the $f(R)$
gravity part of the action actually produces the dark energy
oscillations at redshifts $z\sim 4$.
\end{abstract}


\maketitle

\section{Introduction}

The dark sector of the Universe constitutes the most mysterious
problems in theoretical physics and cosmology, since these two
sectors control the evolution of the Universe to an 96$\%$ extent.
The dark sector consists of two parts, the dark matter and dark
energy part, and both still urge for a consistent explanation.
With regard to dark matter, it is still a question whether it is
controlled by a weakly interactive massive particle
\cite{Bertone:2004pz,Bergstrom:2000pn,Mambrini:2015sia,Profumo:2013yn,Hooper:2007qk,Oikonomou:2006mh},
or it is simply some manifestation of modified version of general
relativity \cite{Capozziello:2012ie}. On the other hand, dark
energy is the name with which the late-time acceleration of the
Universe, firstly observed in the late 90's \cite{Riess:1998cb},
is now known, and this mysterious dark energy era has attracted a
lot of attention in the literature
\cite{Bamba:2012cp,Peebles:2002gy,Li:2011sd,Bamba:2010wb,Frieman:2008sn,Boehmer:2008av,Nojiri:2006gh,Elizalde:2004mq,Makarenko:2018blx,Capozziello:2003gx,Kamenshchik:2001cp,Carroll:1998zi,Capozziello:2002rd,Capozziello:2005ra}.
In all the theoretical approaches towards consistently describing
the dark energy era, modified gravity is to date the most
promising description, see for example the reviews
\cite{reviews1,reviews2,reviews3,reviews4,reviews5,reviews6}.

Apart from the mysterious dark sector of the Universe, another
major issue which hopefully in the next two decades will be
explained, is the primordial post-quantum gravity era of our
Universe. To date there are two candidate descriptions for this
primordial era, the inflationary scenario
\cite{Guth:1980zm,Linde:1993cn,Linde:1983gd} and the bouncing
cosmology scenario
\cite{Brandenberger:2016vhg,deHaro:2015wda,Cai:2014bea,Avelino:2012ue,Koehn:2013upa,Cai:2013kja,Brandenberger:2012zb,Cai:2011zx,Allen:2004vz}.
Both descriptions produce a nearly scale-invariant power spectrum
for the primordial scalar curvature perturbations, with the
cosmological bounces having the attribute of also producing a
cosmological evolution free from the initial singularity.

In the context of modified gravity it is often possible to
describe in an unified way the early and late-time eras of our
Universe, see for example \cite{Nojiri:2003ft,Odintsov:2020nwm}.
In fact, modified gravity might serve as the only consistent
description of dark energy beyond general relativity. The reason
for this is simple, since general relativity can describe
late-time acceleration in a restricted way, by using a scalar
field which produces either quintessential or phantom evolution,
and also a simple cosmological constant may describe a de Sitter
evolution at late-times. But phantom scalar fields are not
necessarily the best description for the late-time era, since
these inevitably drive the Universe towards a Big Rip singularity
\cite{Caldwell:2003vq}, and also phantom fields can be sources of
instabilities. Modified gravity can successfully provide a
consistent late-time era, that can mimic a quintessential or de
Sitter or even a phantom dark energy era, see for example the
reviews
\cite{reviews1,reviews2,reviews3,reviews4,reviews5,reviews6} for
more details on these issues.

One promising sector of modified gravity theories, is the
Gauss-Bonnet gravity
\cite{Li:2007jm,Nojiri:2005jg,Elizalde:2020zcb,Cognola:2006eg,Elizalde:2010jx,Izumi:2014loa,Oikonomou:2016rrv,Kleidis:2017ftt,Oikonomou:2015qha,Escofet:2015gpa,new2,Makarenko:2016jsy,Navo:2020eqt,Bajardi:2020osh,Capozziello:2019wfi,Benetti:2018zhv},
in the context of which, the Gauss-Bonnet invariant appears in the
Lagrangian in a non-linear way. Also, extensions of general
relativity which contain higher orders of the Riemann and Ricci
tensors can be found in Refs.
\cite{Clifton:2006kc,Bogdanos:2009tn,Capozziello:2004us}. The
focus in this paper is in general on $f(R,\mathcal{G})$ theories
of gravity
\cite{Elizalde:2010jx,Bamba:2009uf,DeLaurentis:2015fea,Benetti:2018zhv,delaCruzDombriz:2011wn},
and specifically we mainly focus on theories of the form
$R+f(\mathcal{G})$, in order to avoid having primordial
superluminal perturbation modes, but for reasons that will be
explained shortly, we also study the late-time behavior of an
$f(R)+g(\mathcal{G})$ model. Our aim is two fold: firstly to find
appropriate model of Gauss-Bonnet gravity that may describe in a
consistent way the dark energy era, and secondly to demonstrate
that in some Gauss-Bonnet models it is possible to provide a
unified description of the primordial post-quantum era and the
dark energy era with the same model. One of the models which we
shall present in this paper, is capable of describing primordially
a Type IV singular bounce, and at late-times a dark energy era,
which is mimics the $\Lambda$-Cold-Dark-Matter ($\Lambda$CDM)
model and produces values for the cosmological quantities of
interest that are compatible with the Planck 2018 data on
cosmological parameters \cite{Aghanim:2018eyx}. With regard to the
primordial Type IV singular bounce, the singularity is Type IV
type, so it is a smooth singularity which does not affect the
evolution of the Universe in an extreme way, such as the crushing
types of singularities. In addition, this particular singular
bounce was shown in an earlier work that it generates a nearly
scale invariant power spectrum of the primordial scalar curvature
perturbations, compatible with the latest Planck 2018 constraints
on inflation. Apart from the fact that our Gauss-Bonnet model of
the form $R+f(\mathcal{G})$, can both produce a singular bounce
primordially and a dark energy era compatible with the
$\Lambda$CDM model and the latest Planck observations
\cite{Aghanim:2018eyx}, one major outcome of our work is that this
specific type of models produce a dark energy era free from dark
energy oscillations, known to be present in $f(R)$ gravity models
at large redshifts $z\sim 4$. In fact, in order to verify this
issue, we also examined the late-time phenomenology of a model of
the form $f(R)+g(\mathcal{G})$, and as we demonstrate, this type
of models can also be compatible with both the Planck 2018
observations and the $\Lambda$CDM model, but it is not free from
dark energy oscillations. Thus our work indicates the fact that
the dark energy oscillations are possibly due to the $f(R)$
gravity sector.

\section{Modifying the Gauss-Bonnet Gravity Theoretical Framework for the Dark Energy Era Study}

The focus in this work is, as we already mentioned, the
unification of a singular bounce with the dark energy era, and to
our knowledge this is the first time that this proposal is
quantitatively materialized. In this section we shall present the
theoretical framework of a general $f(R,\mathcal{G})$ gravity and
we shall appropriately modify the Friedmann equation by using
appropriate statefinder quantities, in order we study in an
optimal way the late-time era. The starting point of our work is
obviously the gravitational action and we shall assume an
$f(R,\mathcal{G})$ model accompanied by the presence of perfect
matter fluids, with the following gravitational action,
\begin{equation}
\centering
\label{action}
S=\int{d^4x\sqrt{-g}\left(\frac{f(R,\mathcal{G})}{2\kappa^2}+\mathcal{L}_{(m)}\right)}\, ,
\end{equation}
with $R$ being the Ricci scalar, $\kappa=\frac{1}{M_P}$ is the
gravitational constant, where $M_P$ denotes the reduced Planck
mass, and $\mathcal{G}$ signifies the Gauss-Bonnet invariant
defined as
$\mathcal{G}=R^2-4R_{\alpha\beta}R^{\alpha\beta}+R_{\alpha\beta\gamma\delta}R^{\alpha\beta\gamma\delta}$
with $R_{\alpha\beta}$ and $R_{\alpha\beta\gamma\delta}$ being the
Ricci and Riemann tensor respectively. Lastly, $\mathcal{L}_{(m)}$
is the Lagrangian density of the perfect matter fluids, which
contains all the information for non-relativistic matter, that is
Cold Dark-Matter (CDM) and relativistic matter, so radiation.
Furthermore, we shall assume that the cosmological background
corresponds to that of a flat Friedman-Robertson-Walker (FRW)
metric, with the line element being,
\begin{equation}
\centering
\label{metric}
ds^2=-dt^2+a^2(t)\delta_{ij}dx^idx^j\, ,
\end{equation}
where $a(t)$ denotes the scale factor. As a result, the Ricci and
Gauss-Bonnet scalar are reduced to simpler forms, which read,
\begin{equation}
\centering
\label{R1}
R=12H^2+6\dot H\,  ,
\end{equation}
\begin{equation}
\centering
\label{G}
\mathcal{G}=24H^2(\dot H+H^2)\, ,
\end{equation}
where $H$ signifies Hubble's parameter defined as $H=\frac{\dot
a}{a}$ and as usual, and the ``dot'' implies differentiation with
respect to cosmic time $t$. Thus, by varying the gravitational
action (\ref{action}) with respect to the metric tensor
$g^{\mu\nu}$, the gravitational field equation is derived. Here,
we shall separate our equations in space and time components,
hence the equations of motion are,
\begin{equation}
\centering
\label{motion1}
3f_RH^2=\kappa^2\rho_{(m)}+ \frac{f_RR+f_\mathcal{G}\mathcal{G}-f-6H\dot f_R-24H^3\dot f_\mathcal{G}}{2}\, ,
\end{equation}
\begin{equation}
\centering
\label{motion2}
-2f_R\dot H=\kappa^2(\rho_{(m)}+P_{(m)})+\ddot f_R-H\dot f_R-4H^3\dot f_\mathcal{G}+8H\dot H\dot f_\mathcal{G}+4H^2\ddot f_\mathcal{G}\, ,
\end{equation}
where for simplicity, we denote differentiation with respect to a
scalar function with a subscript. Furthermore, as stated before,
the matter density is comprised of both relativistic and
non-relativistic particles and consequently is written as,
\begin{equation}
\centering
\label{density1}
\rho_{(m)}=\rho_{d0}\left(\frac{1}{a^3(t)}+\chi\frac{1}{a^4(t)}\right)\, ,
\end{equation}
where $\rho_{d0}$ signifies the current value of the
non-relativistic density and $\chi=\frac{\rho_{r0}}{\rho_{d0}}$ is
the ratio of the current values of relativistic and
non-relativistic matter. In addition, $P_{(m)}$ denotes the
corresponding pressure which is connected to the matter density
as,
\begin{equation}
\centering
P_{i}=\omega_i\rho_{i}\, ,
\end{equation}
\begin{equation}
\centering
\label{Pm}
P_{(m)}=\sum_{i}{P_i}\, ,
\end{equation}
with $\omega_i$ being the equation of state parameter for a
specific kind of matter and $i=d, r$, either non-relativistic or
relativistic matter perfect fluids. Both kinds are treated as a
barotropic perfect fluid with continuity equations,
\begin{equation}
\centering
\label{conteq}
\dot\rho_i+3H(\rho_i+P_i)=0\, ,
\end{equation}
\begin{figure}
\centering
\includegraphics[width=20pc]{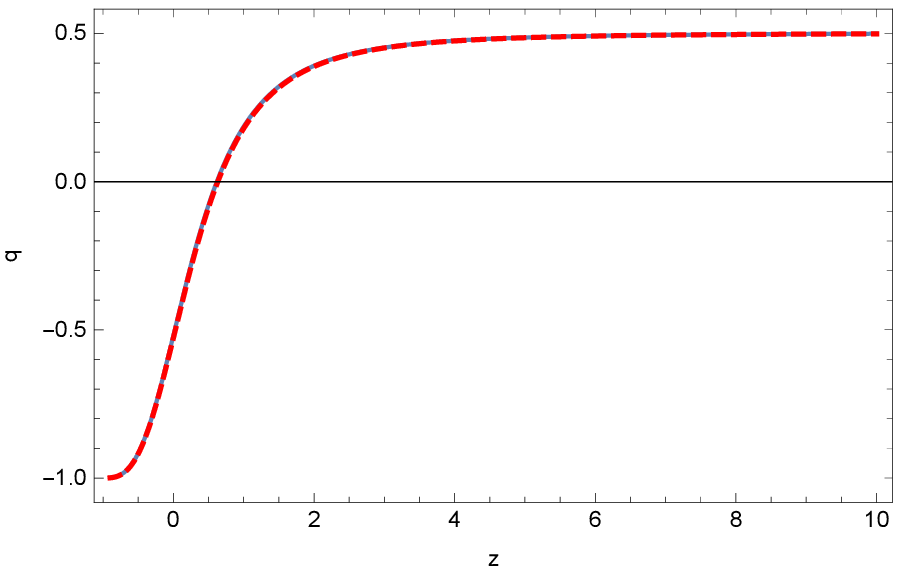}
\includegraphics[width=19.5pc]{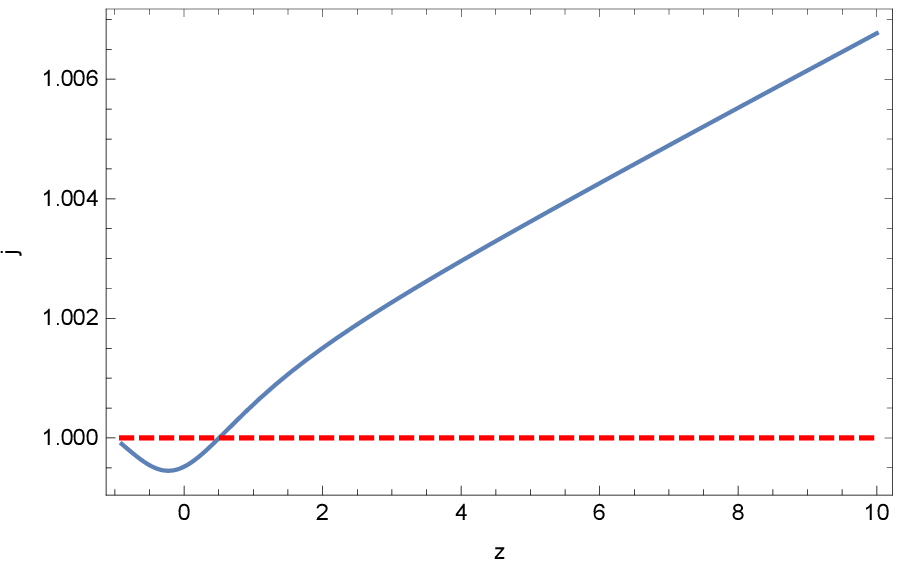}
\includegraphics[width=20pc]{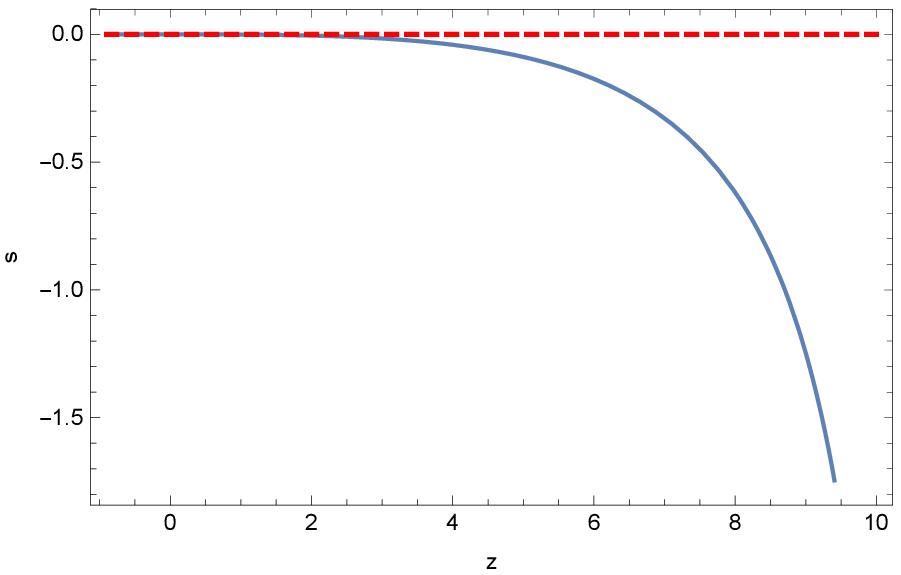}
\includegraphics[width=20pc]{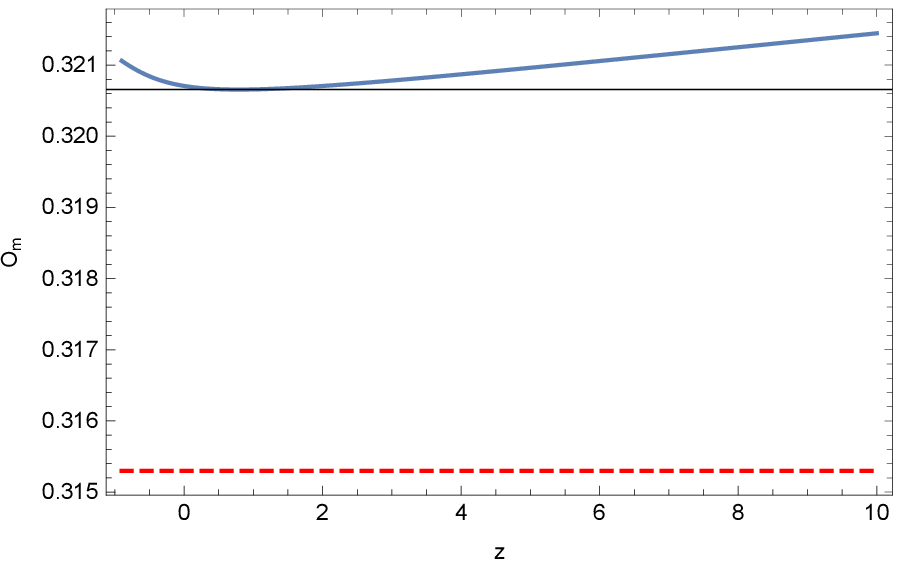}
\caption{Comparison of several statefinder quantities for the
$R+f(\mathcal{G})$ gravity and the $\Lambda$CDM model.}
\label{plot1}
\end{figure}
The aim of our study is to derive a functional form for Hubble's
parameter, hence only a single equation of motion is necessary. In
the following, we shall utilize Eq. (\ref{motion1}) which we aim
to solve numerically, and to extract the form of Hubble rate
during the dark energy era. Before we continue however, we shall
perform certain changes which will facilitate our study.
Specifically, as a dynamical variable we shall use the redshift,
and we shall also introduce a statefinder variable $y_H(z)$ which
we define shortly, in order to make the late-time study more
concrete and easy to tackle numerically.

So the cosmic time will be replaced by a more convenient variable
which is the redshift $z$. From the definition of the redshift,
\begin{equation}
\centering
\label{redshift}
1+z=\frac{1}{a(t)}\, ,
\end{equation}
where we assumed that at present time the scale factor is equal to
unity, a new differential operator can be constructed by simply
performing a differentiation on this particular relation, which in
turn reads,
\begin{equation}
\centering
\label{operator}
\frac{d}{dt}=-H(1+z)\frac{d}{dz}\, ,
\end{equation}
where now, Hubble's parameter depends solely on the redshift, that
is $H=H(z)$. This operator is of paramount importance as each
object in the equations of motion which is differentiated with
respect to the cosmic time, can be transformed to a
redshift-dependent quantity by using the above transformation.
Below we quote some important quantities that will be used
frequently in this paper, and these are transformed as,
\begin{equation}
\centering
\dot H=-H(1+z)H'\, ,
\end{equation}
\begin{equation}
\centering
\dot R=6H(1+z)^2\left(H'^2+HH''-\frac{3HH'}{1+z}\right)\, ,
\end{equation}
\begin{equation}
\centering
\mathcal{\dot G}=24(1+z)^2H^3\left(3H'^2+HH''-\frac{3HH'}{1+z}\right)\, ,
\end{equation}
where the ``prime'' implies differentiation with respect to the
redshift.  Also we have,
\begin{figure}[h!]
\centering
\includegraphics[width=20pc]{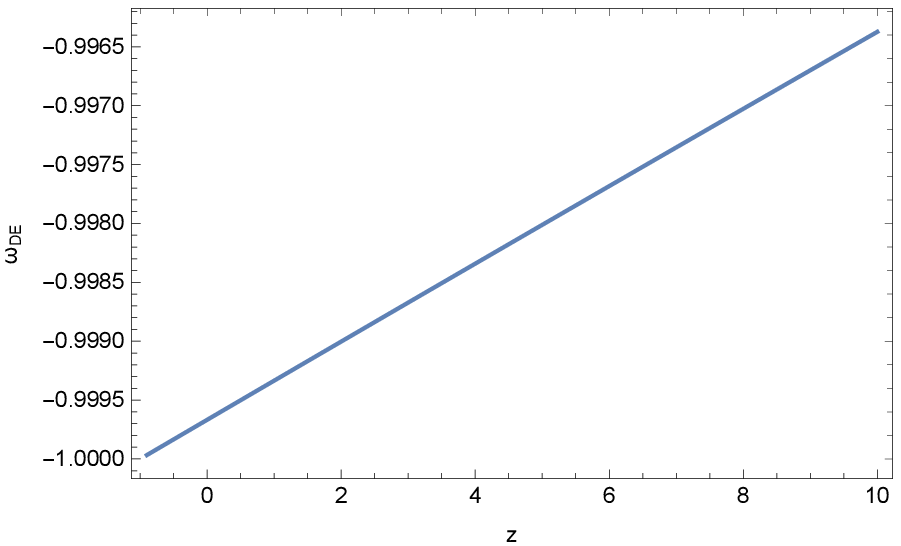}
\includegraphics[width=19pc]{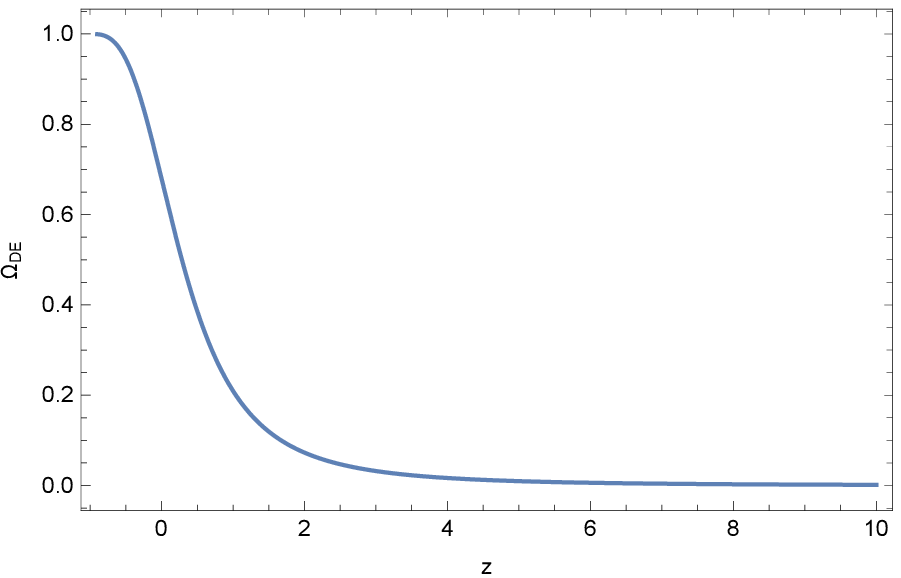}
\caption{The dark energy EoS parameter $\omega_{DE}$ (left) and
dark energy density parameter $\Omega_{DE}$ (right) for the
$R+f(\mathcal{G})$ model. In this case, it can easily be seen that
the EoS of dark energy density parameter is slowly varying near
$\omega_{DE}=-1$ while the density parameter $\Omega_{DE}$
increases until it reaches the value $\Omega_{DE}=1$.}
\label{plot2}
\end{figure}
\begin{align}
\centering
\dot f_X&=-H(1+z)f_X'&\dot f_X&=\sum_{Y}{\dot Yf_{XY}}\, ,
\end{align}
where $X, Y$ take the values $R, \mathcal{G}$. This is because,
\begin{equation}
\centering
\frac{d}{dt}=\frac{dR}{dt}\frac{\partial}{\partial R}+\frac{d\mathcal{G}}{dt}\frac{\partial}{\partial\mathcal{G}}\, ,
\end{equation}
since $R=R(t)$, $\mathcal{G}=\mathcal{G}(t)$ and
$f=f(R,\mathcal{G})$. Both approaches are valid so the choice is
up to the reader, however, even for the first case, a similar
relation for the differential operator with respect to redshift
applies.

The second change which shall be made is a function replacement,
and specifically, instead of using the Hubble rate, we shall use
an appropriate statefinder function related to it. But before we
continue, it is worth making certain changes in the equations of
motion. Recalling Eq. (\ref{motion1}), we shall treat each
geometric term derived from the expression $f(\mathcal{G})$ in the
gravitational action (\ref{action}) as a fluid, corresponding to
dark energy, which will turn out to be a perfect fluid as well.
Assuming that,
\begin{equation}
\centering
\label{DEdensity}
\rho_{DE}=\frac{f_RR+f_\mathcal{G}\mathcal{G}-f-6H\dot f_R-24H^3\dot f_\mathcal{G}}{2\kappa^2}+\frac{3H^2}{\kappa^2}(1-f_R)\, ,
\end{equation}
\begin{equation}
\centering \label{DEpressure} P_{DE}=\frac{\ddot f_R+2H\dot
f_R-3H^2(1-f_R)+8H^3\dot f_\mathcal{G}+8H\dot H\dot
f_\mathcal{G}+4H^2\ddot
f_\mathcal{G}}{\kappa^2}+\frac{f-f_RR-f_\mathcal{G}\mathcal{G}}{2\kappa^2}+\frac{2\dot
H}{\kappa^2}(f_R-1)\, ,
\end{equation}
then the continuity equation reads,
\begin{equation}
\centering
\label{DEcontinuity}
\dot\rho_{DE}+3H(\rho_{DE}+P_{DE})=0\, .
\end{equation}
This continuity equation, as mentioned before, implies that the
dark energy fluid is perfect, as is the case with the rest of the
perfect fluids present. Consequently, the equations of motion
obtain the familiar form of Friedman's  and Raychaudhuri's
equations,
\begin{equation}
\centering
\label{motion3}
H^2=\frac{\kappa^2}{3}(\rho_{(m)}+\rho_{DE})\, ,
\end{equation}
\begin{equation}
\centering \label{motion4} \dot
H=-\frac{\kappa^2}{2}(\rho_{(m)}+P_{(m)}+\rho_{DE}+P_{DE})\, .
\end{equation}
With these equations at hand, we define the new statefinder
function $y_H(z)$ as,
\begin{equation}
\centering
y_H(z)=\frac{\rho_{DE}}{\rho_{d0}}\, .
\end{equation}
This is the new dimensionless function which will participate in
the equations of motion instead of Hubble's parameter. Since the
dark energy density was defined in Eq. (\ref{DEdensity}), this
particular function is related to the Hubble rate, as follows,
\begin{equation}
\centering
\label{yH}
y_H(z)=\frac{H^2}{m_s^2}-\frac{\rho_{(m)}}{\rho_{d0}}\, ,
\end{equation}
where $m_s$ is a mass scale defined as
$m_s^2=\frac{\kappa^2\rho_{d0}}{3}=H_0^2\Omega_{(m)}^{0}$ with
$H_0$ being the current value of Hubble's parameter and
$\Omega_{(m)}^{0}$ the current value of the matter density
parameter. Their values will be assumed to be equal to
$H_0=67.4\pm0.5\frac{km}{sec\times Mpc}$ and
$\Omega_{(m)}^{0}=0.3153$ which are both based on the latest
Planck 2018 data \cite{Aghanim:2018eyx}. This statefinder function
will be used as a replacement for Hubble and its derivatives, as
shown below,
\begin{equation}
\centering
\label{H}
H^2=m_s^2\left(y_H+\frac{\rho_{(m)}}{\rho_{d0}}\right)\, ,
\end{equation}
\begin{equation}
\centering
\label{H'}
HH'=\frac{m_s^2}{2}\left(y_H'+\frac{\rho_{(m)}'}{\rho_{d0}}\right)\, ,
\end{equation}
\begin{equation}
\centering
\label{H''}
H'^2+HH''=\frac{m_s^2}{2}\left(y_H''+\frac{\rho_{(m)}''}{\rho_{d0}}\right)\, ,
\end{equation}
An observant reader might notice that Hubble's derivatives
participate in the previous equations with these exact forms.
Indeed the Ricci scalar and Gauss-Bonnet invariant time derivative
contain the above expressions, so with this designation, all the
previous equations can be rewritten easily. Furthermore, recalling
equations (\ref{density1}) and (\ref{redshift}), the density of
matter is rewritten with respect to the redshift variable as
follows,
\begin{equation}
\centering
\label{density2}
\rho_{(m)}=\rho_{d0}((1+z)^3+\chi(1+z)^4)\, .
\end{equation}
where $\chi\simeq3.1\cdot10^{-4}$, and with $\chi$ being the
fraction of the present day energy density of the radiation over
the cold dark matter fluids. Finally, we define the following
parameters which we shall evaluate during the late-time era.
Concerning the dark energy Equation of State (EoS) parameter, this
is equal to,
\begin{equation}
\centering
\label{DEEoS}
\omega_{DE}=-1+\frac{1+z}{3}\frac{d\ln{y_H}}{dz}\, ,
\end{equation}
while the dark energy density parameter is,
\begin{equation}
\centering \label{DEDenPar}
\Omega_{DE}=\frac{y_H}{y_H+\frac{\rho_{(m)}}{\rho_{d0}}}=y_H\left(\frac{m_s}{H}\right)^2\,
.
\end{equation}
Thus the aim in this paper, is to study some appropriate
$f(R,\mathcal{G})$ models and compare their behavior with the
$\Lambda$CDM model. In order to extract the late-time behavior of
each $f(R,\mathcal{G})$ model chosen, we shall solve numerically
the differential equation (\ref{motion1}) with respect to the
statefinder $y_H(z)$, and for appropriately chosen physically
motivated initial conditions.

\section{$R+f(\mathcal{G})$ Gravity: A Singular Bounce at Early-times and a Dark Energy Era at Late-times}

For an arbitrary $f(R,\mathcal{G})$, the possibility of ghosts
being present is nonzero. For the first model, we shall assume a
ghost free case where function $f(R,\mathcal{G})$ is replaced by
$R+f(\mathcal{G})$. As a result, equation Eq. (\ref{motion1}) is
rewritten as

\begin{equation}
\centering
\label{}
\frac{3H^2}{\kappa^2}=\rho_{(m)}+\frac{\mathcal{G}f_{\mathcal{G}}-f-24\dot f_{\mathcal{G}}H^3}{2\kappa^2}\, ,
\end{equation}
This is the general differential equation that must be solved in
the interval [-0.9,10] for redshift with respect to $y_H(z)$
defined in Eq. (\ref{yH}). Let us assume now that the Gauss-Bonnet
function is given by the following expression
\begin{equation}
\centering \label{F1}
f(\mathcal{G})=\frac{c_1}{\mathcal{G}}+c_2\mathcal{G}^{\frac{\alpha}{3\alpha-1}}\,
,
\end{equation}
where $c_1$ and $c_2$ are auxiliary constants with mass dimensions
$[m]^{6}$ and $[m]^{2-\frac{4\alpha}{3\alpha-1}}$ respectively for
consistency while $\alpha$ is an additional parameter which we
shall assume it satisfies the condition $\alpha>1$.
\begin{figure}[h!]
\centering
\includegraphics[width=20pc]{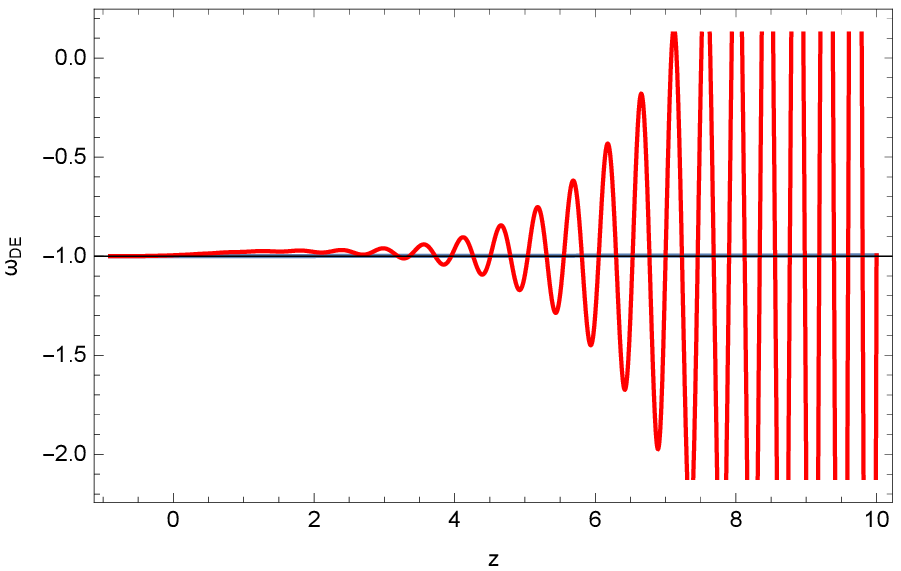}
\includegraphics[width=19pc]{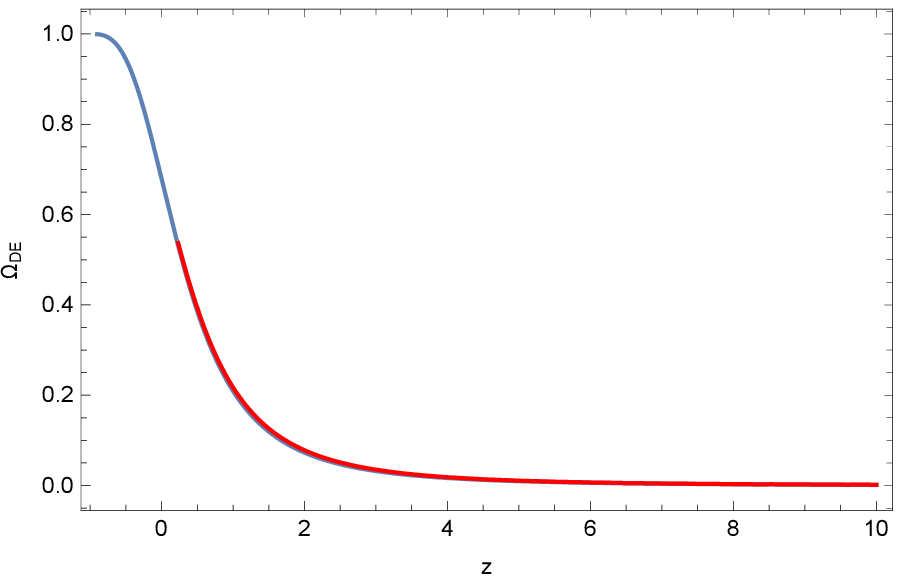}
\caption{Dark energy EoS parameter $\omega_{DE}$ (left) and the
dark energy density parameter $\Omega_{DE}$ (right) for the $f(R)$
and $R+f(\mathcal{G})$ cases, depicted with red and blue curves
respectively. } \label{plot3}
\end{figure}
This is an interesting model due to the fact that for small values
of $\mathcal{G}$, the inverse term $\frac{c_1}{\mathcal{G}}$
becomes dominant during the late time whereas
$c_2\mathcal{G}^{\frac{\alpha}{3\alpha-1}}$ is dominant during
the early time where the exponent $\frac{\alpha}{3\alpha-1}$ is
lesser than unity given that the condition $\alpha>1$ holds true.
Therefore, this model is capable of describing both early and late
time era and thus unifying them smoothly.

The singular bounce cosmology is basically realized by an
$R+f(\mathcal{G})$ gravity of the form $f(\mathcal{G})\sim
\mathcal{G}^{\frac{1-2\alpha}{3\alpha-1}}$ as it was shown in Ref.
\cite{Oikonomou:2015qha}. This particular $f(\mathcal{G})$ gravity
was able to realize a Type IV singular bounce with Hubble rate
$H(t)\sim \left(t-t_s\right)^{\alpha }$, with $\alpha$ strictly
greater than unity, that is $\alpha>1$, and $t_s$ is the cosmic
time instance that the singular bounce occurs. From Eq. (\ref{F1})
it is apparent that for $\alpha>1$ the term $\sim
\mathcal{G}^{\frac{\alpha}{3\alpha-1}}$ is dominant at early
times, and the term $\sim G^{-1}$ is subdominant during the
early-time era. We shall quantify this shortly, but let us discuss
in short the singular bounce generation by the term $\sim
\mathcal{G}^{\frac{\alpha}{3\alpha-1}}$. Following Ref.
\cite{Oikonomou:2015qha}, the primordial curvature perturbations
are generated near the bouncing point $t=t_s$, and exit the Hubble
horizon after the singular bouncing point. The Hubble rate near
the bouncing point could be of the order $H_I\sim 10^{13}$GeV
(borrowing the value of the Hubble rate from low-scale inflation
studies), thus, the term $\sim
\mathcal{G}^{\frac{\alpha}{3\alpha-1}}$ for $\alpha=3.30579$ and
$c_2=1$eV$^{2-\frac{4\alpha}{3\alpha-1}}$, which are the values of
the parameters $\alpha$ and $c_2$ we shall use in the following,
is of the order $c_2 \mathcal{G}^{\frac{\alpha}{3\alpha-1}}\sim
1.11887\times 10^{30}$eV$^2$, while the term $c_1\mathcal{G}^{-1}$
is of the order $c_1\mathcal{G}^{-1}\sim 10^{-88}$eV$^2$, for
$c_1=1$eV$^6$. Thus indeed, the term $\sim G^{-1}$ is
significantly subdominant during the early-time era, near the
singular bouncing point. In Ref. \cite{Oikonomou:2015qha}, we
calculated in detail the power spectrum of the primordial scalar
curvature perturbations, and it was found that it is equal to,
\begin{equation}\label{powerspectrumfinal}
\mathcal{P}_R\sim k^{\frac{7}{2}+3+\frac{\left(2-2 \alpha +\alpha
^2\right) \mu }{2 (-1+\alpha )}}\, .
\end{equation}
From the above expression we easily derived the spectral index of
the primordial curvature perturbations, which is equal to,
\begin{equation}
 n_s-1\equiv\frac{d\ln\mathcal{P}_{\mathcal{R}}}{d\ln
k}=\frac{7}{2}+3+\frac{2-2\alpha+\alpha^2}{2(\alpha-1)}\mu=1-\frac{11}{2(\alpha-
1)^2}~.
\end{equation}
It is easy to see that for $\alpha$ taken in the range
$\alpha=[3.2999,3.31]$, the spectral index becomes compatible with
the latest Planck 2018 data \cite{Akrami:2018odb}, which constrain
the spectral index to be $n_s=0.9649\pm 0.0042$. Also for
$\alpha<-1$ the spectral index can also be compatible with the
Planck data, however the $\alpha<-1$ case corresponds to a Big Rip
singularity, thus it is not physically acceptable. The singular
bounce which has a Type IV singularity, is more physically
appealing, since the Type IV singularity is quite smooth and does
not affect any physical quantities that can be defined on the
three dimensional spacelike hypersurface which is defined at the
time instance that the singularity occurs.

Having settled that the term practically generates the singular
bounce at early times, and is dominant during this primordial era,
let us see how things are modified at late times. Apparently, the
late-time era is controlled by the term $\sim \mathcal{G}^{-1}$,
but let us see this explicitly in a quantitative way for the
moment. Let us use the current value of the Hubble rate which is
$H_0\sim 10^{-33}$eV, so the term
$\mathcal{G}^{\frac{\alpha}{3\alpha-1}}$ is approximately of the
order $c_2\mathcal{G}^{\frac{\alpha}{3\alpha-1}}\sim 8.44947\times
10^{-46}$eV$^2$, while the term $c_1\mathcal{G}^{-1}\sim
10^{132}$eV$^2$. Thus it is quantitatively apparent that the term
$\sim \mathcal{G}^{-1}$ is quite dominant at late times and
controls the evolution.

In this section we shall demonstrate numerically that the term
$\mathcal{G}^{-1}$ indeed dominates the late time era, generating
a viable dark energy era. Our results are robust towards changes
of the parameter $\alpha$ for all values larger than unity, but
the analysis will be focused on those values of the parameter
$\alpha$ that yield a spectral index of the primordial curvature
perturbations compatible with the Planck 2018 data, so
$\alpha=3.30579$. So let us proceed with the analysis of the model
and express all the differential equations and the physical
quantities in terms of the statefinder quantity $y_H(z)$. Before
going to the details of our analysis, let us quote some useful
expressions, so for the $f(\mathcal{G})$ gravity chosen as in Eq.
(\ref{F1}), we have,
\begin{equation}
\centering \label{fG1}
f_{\mathcal{G}}=-\frac{c_1}{\mathcal{G}^2}+\frac{c_2\alpha}{3\alpha-1}\mathcal{G}^{\frac{1-2\alpha}{3\alpha-1}}\,
,
\end{equation}

\begin{equation}
\centering \label{fG1dot} \dot f_{\mathcal{G}}=\mathcal{\dot
G}\left(\frac{2c_1}{\mathcal{G}^3}+\frac{c_2\alpha(1-2\alpha)}{(3\alpha-1)^2}\mathcal{G}^{\frac{2-5\alpha}{3\alpha-1}}\right)\,
,
\end{equation}
and thus Eq. (\ref{motion1}) reads,
\begin{equation}
\centering \label{motion5}
3H^2=\kappa^2\rho_{(m)}+\frac{2c_1(1-3\alpha)+c_2(2\alpha-1)\mathcal{G}^{\frac{4\alpha-1}{3\alpha-1}}}{2\mathcal{G}(3\alpha-1)}-12\mathcal{\dot
G}\left(\frac{2c_1}{\mathcal{G}^3}+\frac{c_2\alpha(1-2\alpha)\mathcal{G}^{\frac{2-5\alpha}{3\alpha-1}}}{(3\alpha-1)^2}\right)H^3\,
.
\end{equation}
This particular differential equation shall be solved
numerically\footnote{We used Mathematica 11.3\circledR} for
redshifts in the range $z=[-0.9,10]$, however for the statefinder
function $y_H(z)$ introduced as a replacement for the Hubble in
Eq. (\ref{yH}). With regard to the initial conditions we shall
choose for the statefinder $y_H(z)$, these are
$y_H(z=10)=\frac{\Lambda}{3m_s^2}\left(1+\frac{1+z_f}{100}\right)$
and $y_H'(z=10)=\frac{\Lambda}{3m_s^2}\frac{1}{1000}$ for $z_f=10$
and there is a strong physical motivation for using these initial
conditions, see for example Ref. \cite{Odintsov:2020nwm}. In
addition, the choice for $\Lambda$ is
$\Lambda=1.1895\cdot10^{-66}$eV, while the mass scale $m_s$ is
$m_s=4.32552\cdot10^{-34}$eV and in addition in the following we
shall take $c_1=1$eV$^6$,
$c_2=1$eV$^{2-\frac{4\alpha}{3\alpha-1}}$, $\alpha=3.30579$. Then
by solving numerically the differential equation (\ref{motion5}),
by using the aforementioned initial conditions and values for the
free parameters, we shall analyze several statefinder quantities
of cosmological interest for the late-time era, and we shall
compare the results with the $\Lambda$CDM model and an $f(R)$
gravity model which is known to produce viability for the
late-time era. The comparison of the results of the Gauss-Bonnet
model with the $\Lambda$CDM is obvious, since the latter is the
cornerstone model of late-time phenomenology, since it is highly
compatible with the CMB. However we need to discuss the comparison
of the Gauss-Bonnet model with the $f(R)$ gravity theory, since
these two are apparently two distinct and phenomenologically
competing theories. Our motivation is simply to investigate
whether the dark energy oscillations at large redshifts ($z\geq
4$) persist in the Gauss-Bonnet theory case. Our results are quite
interesting, since in the Gauss-Bonnet case the oscillations do
not occur. Also we shall calculate the predicted values for some
quantities of cosmological interest and compare these values with
the latest constraints of the Planck collaboration on these
cosmological parameters \cite{Aghanim:2018eyx}.
\begin{figure}[h!]
\centering
\includegraphics[width=20pc]{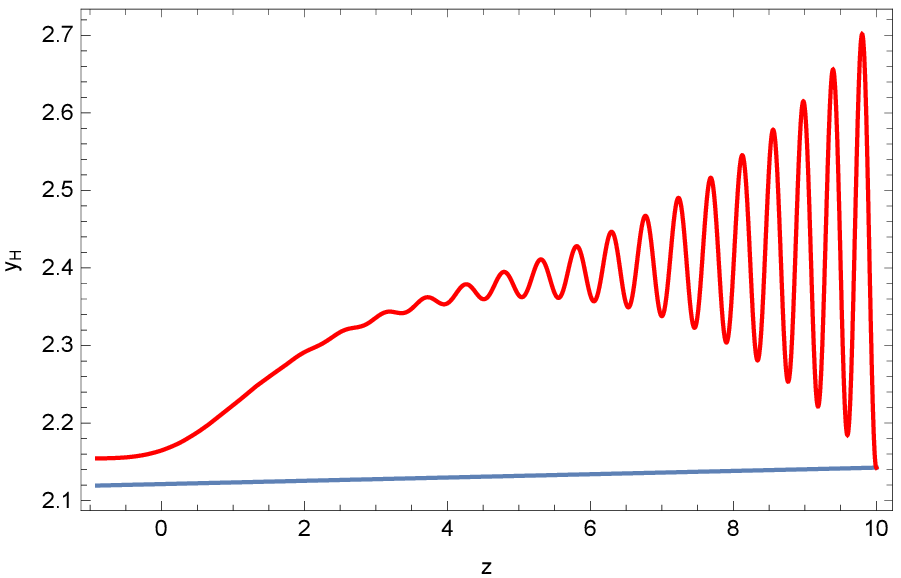}
\includegraphics[width=20pc]{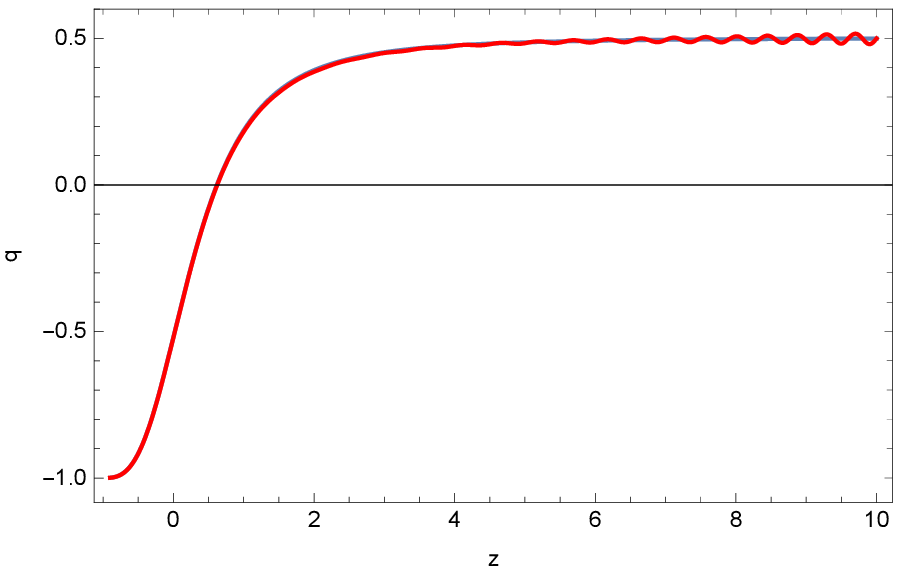}
\caption{Comparison between the $f(R)$ (red) and
$R+f(\mathcal{G})$ case (blue) on parameters $y_H(z)$ (left) and
$q(z)$ (right). While statefinder $y_H$ is completely different
between the two models, the deceleration parameter seems the same
when dark energy oscillations are not present, meaning for
$z\le5$.} \label{plot4}
\end{figure}
For our analysis, we shall use the CMB based value for the Hubble
rate, which is, \cite{Aghanim:2018eyx},
\begin{equation}\label{H0today}
H_0=67.4\pm 0.5 \frac{km}{sec\times Mpc}\, ,
\end{equation}
so $H_0=67.4km/sec/Mpc$ which is $H_0=1.37187\times 10^{-33}$eV.
Also let us discuss in brief the cosmological quantities and
statefinders which we shall analyze and compare their behavior in
this paper, for the $\Lambda$CDM, the Gauss-Bonnet and the $f(R)$
gravity model. An important quantity is the dark energy equation
of state (EoS) parameter, defined as
$\omega_{DE}=\frac{P_{DE}}{\rho_{DE}}$, which in terms of the
statefinder quantity $y_H(z)$ is expressed as follows,
\begin{equation}\label{omegade}
\omega_{DE}(z)=-1+\frac{1}{3}(z+1)\frac{1}{y_H(z)}\frac{d
y_H(z)}{d z}\, .
\end{equation}
Basically, the above quantity is itself a statefinder quantity,
since it depends on the geometry through its explicit dependence
on the Hubble rate derivatives. Another important cosmological
quantity is the dark energy density parameter $\Omega_{DE}(z)$,
defined as $\Omega_{DE}=\frac{\rho_{DE}}{\rho_{tot}}$, which in
terms of the statefinder quantity $y_H(z)$ is written as,
\begin{equation}\label{omegaglarge}
\Omega_{DE}(z)=\frac{y_H(z)}{y_H(z)+(z+1)^3+\chi (z+1)^4}\, .
\end{equation}
Now, for the comparisons with the $\Lambda$CDM model, the Hubble
rate for the $\Lambda$CDM model is equal to,
\begin{equation}\label{lambdacdmhubblerate}
H_{\Lambda}(z)=H_0\sqrt{\Omega_{\Lambda}+\Omega_M(z+1)^3+\Omega_r(1+z)^4}\,
,
\end{equation}
where again $H_0$ is the value of the Hubble rate at present time,
namely, $H_0\simeq 1.37187\times 10^{-33}$eV
\cite{Aghanim:2018eyx}, while $\Omega_{\Lambda}\simeq 0.681369$
and finally $\Omega_M\sim 0.3153$ \cite{Aghanim:2018eyx}. Also
$\Omega_r/\Omega_M\simeq \chi$, where we defined the parameter
$\chi$ below Eq. (\ref{density2}).
\begin{figure}[h!]
\centering
\includegraphics[width=20pc]{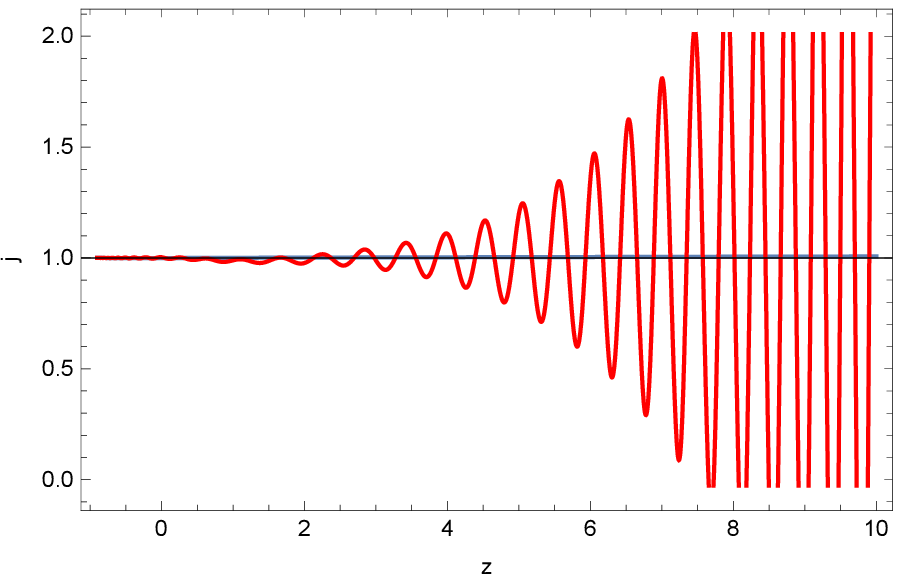}
\includegraphics[width=19.5pc]{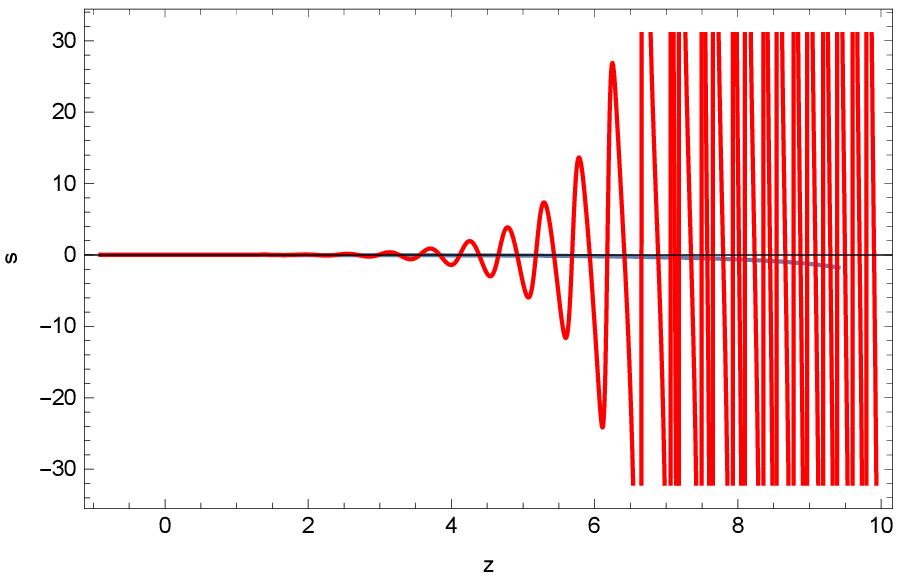}
\includegraphics[width=20pc]{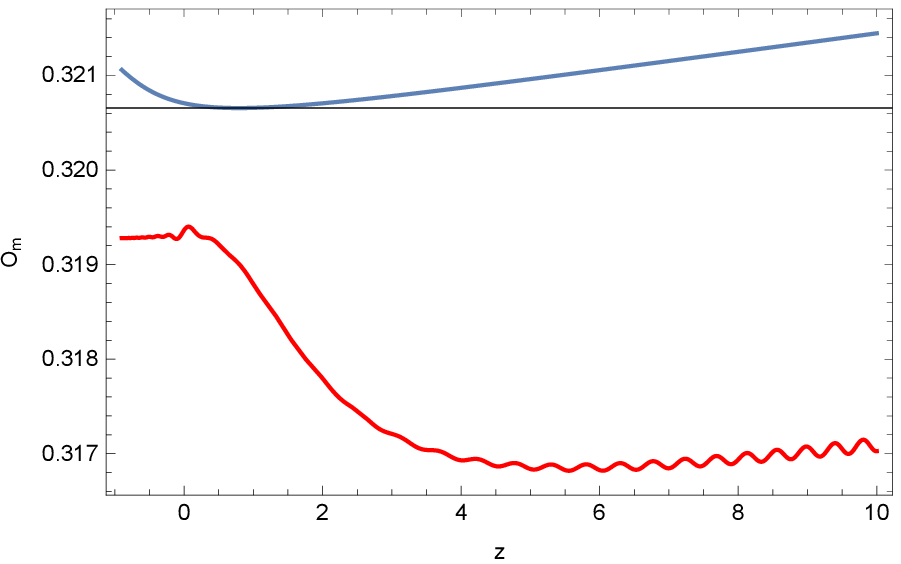}
\caption{Statefinder parameters $j(z)$ (upper left), $s(z)$ (upper
right) and $Om(z)$ (bottom) for $f(R)$ (red) and
$R+f(\mathcal{G})$ (blue) models. In the upper diagrams, it
becomes apparent that the main difference lies in the dark energy
oscillations which as expected is enhanced in higher order
derivatives of $y_H$. } \label{plot5}
\end{figure}

Finally, regarding the statefinder quantities we shall examine in
this late-time study, our focus will be on four statefinder
quantities, the deceleration parameter $q$, the jerk $j$, the snap
parameter $s$, and finally the parameter $Om(z)$, which in terms
of the Hubble rate are defined as follows,
\begin{equation}
\centering \label{q} q=-1-\frac{\dot H}{H^2}\, ,
\end{equation}
\begin{equation}
\centering \label{j} j=\frac{\ddot H}{H^3}-3q-2\, ,
\end{equation}
\begin{equation}
\centering \label{s} s=\frac{j-1}{3(q-\frac{1}{2})}\, ,
\end{equation}
\begin{equation}
\centering \label{Om}
Om(z)=\frac{\left(\frac{H(z)}{H_0}\right)^2-1}{(1+z)^3-1}\, .
\end{equation}
Of course, the first three must be expressed in terms of the
redshift variable, so for simplicity we quote the simplest
expressions of the three, and these are,
\begin{equation}
\centering \label{qz} q=-1+(1+z)\frac{H'}{H}\, ,
\end{equation}
\begin{equation}
\centering \label{jz}
j=(1+z)^2\left(\left(\frac{H'}{H}\right)^2+\frac{H''}{H}-\frac{2H'}{(1+z)H}\right)+1\,
.
\end{equation}
For the $\Lambda$CDM model, the statefinders $j$, $s$ and $Om(z)$
have the following simple values,  $s=0$, $j=1$ and
$Om(z)=\Omega_M\simeq 0.3153$. Let us now proceed to the results
of our numerical analysis. Firstly we shall compare the
Gauss-Bonnet model with the $\Lambda$CDM model, and in Fig.
\ref{plot1} we present the comparisons of the deceleration
parameter (left upper plot), the jerk (right upper plot), the snap
(bottom left plot) and the parameter $Om(z)$ (bottom right plot),
for the Gauss-Bonnet model (blue curves) and the $\Lambda$CDM
model (red curves). As it is obvious, in the case of the
deceleration parameter, the two curves are indistinguishable,
while differences can be found for the rest three statefinders.
Also it is mentionable that for the snap parameter, for redshifts
$z\sim 4$ and smaller, the $\Lambda$CDM model and the Gauss-Bonnet
model are indistinguishable.

Now in order to investigate also the dark energy oscillations
issue, known to affect the $f(R)$ gravity theories, we shall
present the results of our numerical analysis for the Gauss-Bonnet
theory, focusing on the dark energy EoS $\omega_{DE}$ and the dark
energy density parameter $\Omega_{DE}(z)$, the behavior of which
is plotted in Fig. \ref{plot2} for the pure Gauss-Bonnet theory.
As it can be seen in Fig. \ref{plot2}, no dark energy oscillations
occur, but in order to make this result more clear, we shall
compare the Gauss-Bonnet theory with a viable $f(R)$ gravity
theory, the functional form of which is \cite{Odintsov:2020nwm},
\begin{equation}
\centering \label{fR}
f(R)=R+\left(\frac{R}{M}\right)^2-\gamma\Lambda\left(\frac{R}{3m_s^2}\right)^\delta\,
,
\end{equation}
where, $M$ is an auxiliary parameter with mass dimensions $[m]$
and is given by the expression $M=1.5\cdot10^{-5}\frac{50}{N}M_P$,
with $N$ being the $e$-foldings number referring to the
inflationary era ($N=60$) and $M_P$ is the reduced Planck mass.
Essentially, the $R^2$ term contributes to the inflationary era
and early time whereas $\left(\frac{R}{3m_s^2}\right)^\delta$ for
$\delta<1$ becomes dominant in the late-time era. In Fig.
\ref{plot3} we plot the behavior of the dark energy EoS
$\omega_{DE}$ (left) and the dark energy parameter $\Omega_{DE}$
(right) for the $f(R)$ (red curves) and the Gauss-Bonnet gravity
(blue curves). As it is apparent, the behavior of the dark energy
density parameter is indistinguishable between the models,
however, the dark energy EoS for the $f(R)$ gravity model has
strong oscillations for $z\geq 4$, which are absolutely absent
from the Gauss-Bonnet model. Thus the dark energy oscillations
plague that haunted the $f(R)$ gravity models, is absent from the
Gauss-Bonnet models. This claim is further supported by the plots
appearing in Figs. \ref{plot4} and \ref{plot5}. In Fig.
\ref{plot4} we compare the deceleration parameter $q$ and the
statefinder $y_H(z)$ for the $f(R)$ gravity model (red curves) and
the Gauss-Bonnet model (blue). The absence of oscillations in both
cases are obvious, for the Gauss-Bonnet case, and the same
conclusions can be derived if we look at Fig. \ref{plot5} where we
plot the statefinder parameters $j(z)$ (upper left), $s(z)$ (upper
right) and $Om(z)$ (bottom) for the $f(R)$ gravity (red curves)
and the Gauss-Bonnet gravity (blue curves). In the upper diagrams,
it becomes apparent that the main difference lies in the dark
energy oscillations which as expected is enhanced in higher order
derivatives of $y_H$, for the $f(R)$ gravity case.
\begin{figure}[h!]
\centering
\includegraphics[width=20pc]{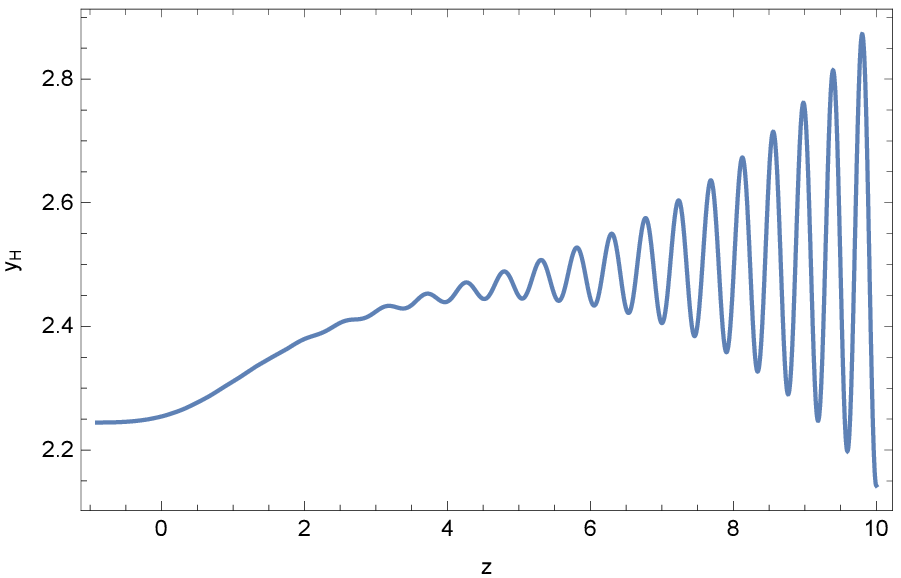}
\includegraphics[width=20pc]{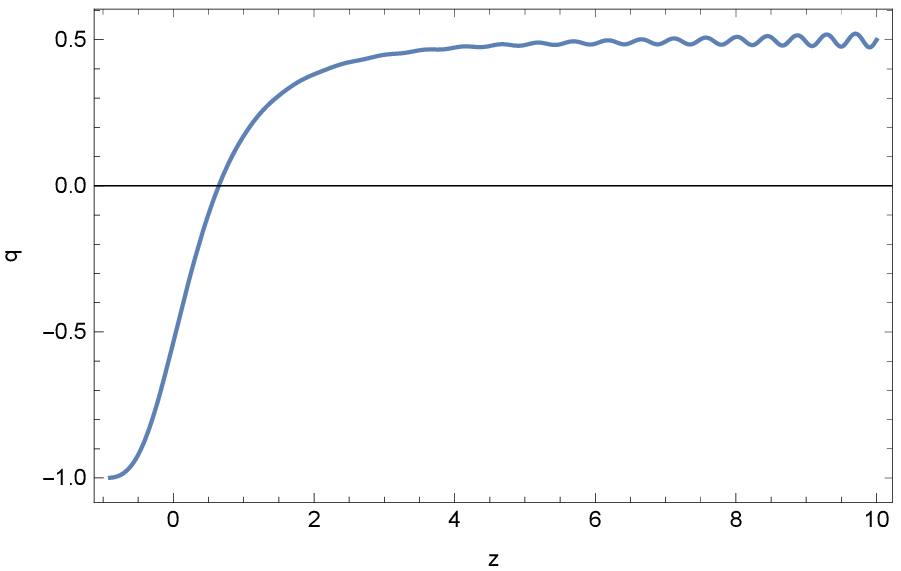}
\caption{Functions $y_H(z)$ (left) and $q(z)$ (right) for the
$f(R)+g(\mathcal{G})$ model with the exact same initial conditions
as in the previous section. In this case, we clearly see that dark
energy oscillations have not been nullified and as a matter of
fact, the same behavior as in the pure $f(R)$ case is produced.
The only difference lies in the numerical values of the
statefinders which shall be addressed shortly.} \label{plot6}
\end{figure}

Finally, in order to have a concrete idea of how well the
Gauss-Bonnet model behaves, we shall compare the values of several
statefinders, for the $\Lambda$CDM model and the Gauss-Bonnet
gravity, and also we directly compare the dark energy EoS and the
dark energy density parameters at present time for the
Gauss-Bonnet model with the latest Planck data. Our results are
summarized in Table \ref{table1}. As it can be seen in Table
\ref{table1}, the statefinders values for the Gauss-Bonnet model
are quite close to the corresponding $\Lambda$CDM value, and in
addition, the Gauss-Bonnet model value for the dark energy density
parameter is $\Omega_{DE}(0)=0.679553$, which is compatible with
the latest Planck constraints $\Omega_{DE}=0.6847\pm 0.0073$
\cite{Aghanim:2018eyx}. In addition, the dark energy EoS parameter
value for the Gauss-Bonnet model at present time is
$\omega_{DE}(0)=-0.999667$ which is in good agreement with the
corresponding Planck constraint $\omega_{DE}=-1.018\pm 0.031$
\cite{Aghanim:2018eyx}.
\begin{table}[h!]
\label{table1} \caption{Cosmological parameters of
$R+f(\mathcal{G})$ choice in the present day versus the expected
value of the $\Lambda$CDM model or the Planck 2018 Data were
available. It becomes apparent that this particular choice for
initial conditions results in values which are quite close to the
expected ones.}
\begin{center}
\begin{tabular}{|r|r|r|}
\hline \textbf{Parameter}&\textbf{Value}&\textbf{$\Lambda$CDM
Value/Planck 2018 Data}\\
\hline $y_H(z=0)$ &2.1213&-
\\ \hline
$y_H'(z=10)$&0.002119&-
\\ \hline q(z=0)&-0.51894&-0.535
\\ \hline
j(z=0)&0.99952&1
\\ \hline
j(z=10)&1.00677&1
\\ \hline
s(z=0)&-0.00015711&0
\\ \hline $Om(z=0)$&0.320707&0.3153 $\pm$
0.07
\\ \hline $\Omega_{DE}(z=0)$&0.679553&0.6847 $\pm$ 0.0073
\\
\hline $\omega_{DE}(z=0)$&-0.999667&-1.018 $\pm$ 0.031
\\ \hline
\end{tabular}
\end{center}
\end{table}
In conclusion, the Gauss-Bonnet model of Eq. (\ref{F1}) is able to
produce a phenomenologically viable early-time bounce, while at
late-times it produces a viable dark energy era, which is in
addition free from dark energy oscillations. In the next section,
we shall also discuss a more complicated $f(R,\mathcal{G})$ model,
however less appealing in comparison to the model we presented in
this section, due to the primordial ghost modes in those
$f(R,\mathcal{G})$ models.
\begin{figure}[h!]
\centering
\includegraphics[width=20pc]{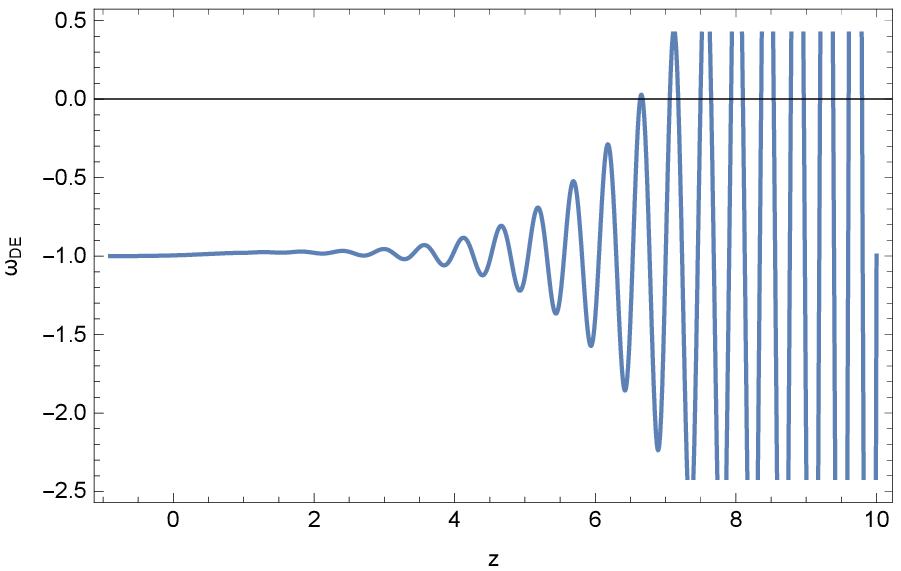}
\includegraphics[width=19pc]{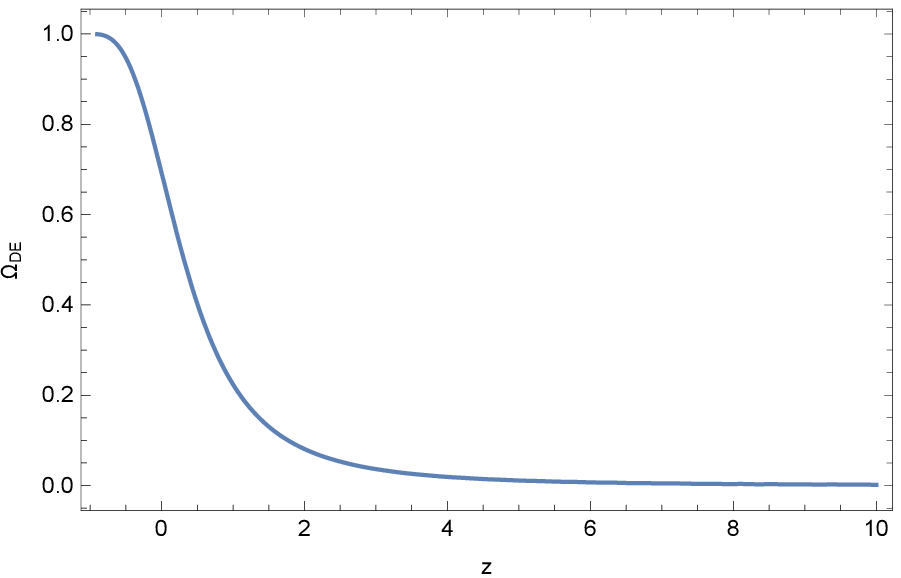}
\caption{The dark energy EoS parameter $\omega_{DE}$ (left) and
the dark energy density parameter $\Omega_{DE}$ (right). As was
the case with the pure $f(R)$ case, only the EoS parameter appears
to oscillate for large redshifts.} \label{plot7}
\end{figure}

\subsection{On the Stability of the $F(G)$ Gravity Solutions and Cosmological Perturbations at the Vicinity of the Bouncing Point}

Let us now investigate the stability of the $f(\mathcal{G})$
gravity solutions and discuss the stability of the cosmological
perturbations as the general relativistic limit is approached. The
FRW equations for the $f(\mathcal{G})$ gravity, constitute a
dynamical system, and the stability of the solutions can be
examined if we perturb this dynamical system. We shall consider
linear perturbations of the cosmological solutions, as functions
of the Hubble rate, and the presence of an instability would
indicate that the solution is not the final attractor for the
theory at hand. For the era near the bouncing point, this is
somewhat expected because the evolution continues after the
bouncing point, but for the late-time era, the expected behavior
is rather vague. This is due to the fact that the late-time era
seems to be a de Sitter like solution by looking at the dark
energy EoS parameter, but eventually it is not an exact de Sitter
solution. So let us check explicitly the stability of the
dynamical system towards linear perturbations of the cosmological
solutions, in order to shed some light on this issue. The Hubble
rate at the vicinity of the bouncing point is,
\begin{equation}\label{hubratepresentpaper}
H(t)=\beta \left(t-t_s\right)^{\alpha }\, ,
\end{equation}
and the $f(\mathcal{G})$ gravity which approximately realizes the
Hubble rate (\ref{hubratepresentpaper}) is,
\begin{equation}\label{fg}
f(\mathcal{G})\sim c_2\mathcal{G}^{\frac{\alpha}{3\alpha-1}}\, .
\end{equation}
Following the strategy of Ref. \cite{Bamba:2014mya}, we linearly
perturb the solution of the Friedmann equation $g(N)=H^2$, in the
following way,
\begin{equation}\label{pert2}
g(N)\rightarrow g(N)+\delta g(N)\, ,
\end{equation}
where $g(N)$ is a solution of the Friedmann equation, and $N$ is
the $e$-foldings number. By using the function $g(N)$, the
Friedmann equation can be cast in the following form,
\begin{align}\label{sfrw12}
& 288g^2(N)f''(\mathcal{G})\Big{[}\left ((g'(N))^2+g(N)\right
)g''(N)+4g(N)g'(N)+4g(N)g'(N)\Big{]}\\ \notag &
6g(N)+f(\mathcal{G})-12g(N)\left (g'(N)+2g(N) \right
)f'(\mathcal{G})=0\, .
\end{align}
Now, the actual conditions which ensure that the dynamical system
of Eq. (\ref{sfrw12}) is stable towards linear perturbations, are
\cite{Bamba:2014mya},
\begin{equation}\label{stcondgg}
\frac{J_2}{J_1}>0,{\,}{\,}{\,}\frac{J_3}{J_1}>0\, ,
\end{equation}
where $J_1$ stands for,
\begin{align}\label{st11}
J_1=288 g(N)^3 f''(\mathcal{G})\, ,
\end{align}
while $J_2$ is,
\begin{align}\label{st12}
J_2=432 g(N)^{2 }\Big{(}(2 g(N)+g'(N)) f''(\mathcal{G})+8 g(N)
\Big{(}g'(N)^2+g(N) (4 g'(N)+g''(N))\Big{)}
f''(\mathcal{G})\Big{)}\, ,
\end{align}
and finally $J_3$ is,
\begin{align}\label{st13}
& J_3=6 \Big{(}1+24 g(N)\Big{(}-8 g(N)^2+3 g'(N)^2+6 g(N) (3
g'(N)+g''(N))\Big{)}f''(\mathcal{G})
\\ \notag & +24 g(N) (4 g(N)+g'(N)) \Big{(}g'(N)^2+g(N) (4 g(N)+g''(N))\Big{)}f''(\mathcal{G})\Big{)}\, .
\end{align}
Now let us calculate explicitly the parameters $J_1$, $J_2$ and
$J_3$ for the era corresponding to the bouncing point, so the
$f(\mathcal{G})$ gravity is given in Eq. (\ref{fg}). By expressing
the Hubble rate (\ref{hubratepresentpaper}) as a function of the
$e$-foldings number $N$, with the latter being equal $N=\ln a$,
the function $g(N)$ becomes,
\begin{equation}\label{gnfunctioform}
g(N)=\frac{\beta ^2 N^{\zeta }}{\delta}\, .
\end{equation}
with $\zeta=\frac{2\alpha}{1+\alpha}$. Now let us proceed with the
stability conditions, and by calculating the parameters $J_1$,
$J_2$ and $J_3$, we get,
\begin{align}\label{st17}
\frac{J_2}{J_1}=\frac{3 \left(\delta^2 N (2 N+\zeta )+8 N^{2 \zeta
} \beta ^4 \zeta  (-1+4 N+2 \zeta )\right)}{2 \delta^2 N^2}\,
\end{align}
and $J_3/J_1$,
\begin{align}\label{st18}
&\frac{J_3}{J_1}=-\frac{1}{c_2 \delta \alpha  (-1+2 \alpha
)}3^{1+\frac{\alpha }{1-3 \alpha }} 4^{\frac{\alpha }{1-3 \alpha
}} N^{-2+\zeta } (1-3 \alpha )^2 \beta ^2 (\zeta )^2
\left(\frac{N^{-1+2 \zeta } \beta ^4 (\zeta
)}{\delta^2}\right)^{\frac{\alpha }{1-3 \alpha }}\times \\ \notag
& \Big{(}1+\frac{2^{\frac{1-\alpha }{-1+3 \alpha }} 3^{\frac{1-2
\alpha }{-1+3 \alpha }} c_2 \delta N^{-\zeta } \alpha  (-1+2
\alpha ) \left(\frac{N^{-1+2 \zeta } \beta ^4 (2 N+\zeta
)}{\delta^2}\right)^{\frac{\alpha }{-1+3 \alpha }} \left(8 N^2-18
N \zeta +3 (2-3 \zeta ) \zeta \right)}{(1-3 \alpha )^2 \beta ^2 (2
N+\zeta )^2}\\ \notag & -\frac{2^{\frac{1-\alpha }{-1+3 \alpha }}
3^{\frac{1-2 \alpha }{-1+3 \alpha }} c_2 \alpha  (-1+2 \alpha )
\left(\frac{N^{-1+2 \zeta } \beta ^4 (\zeta
)}{\delta^2}\right)^{\frac{\alpha }{-1+3 \alpha }} (\zeta ) (\zeta
(-1+2 \zeta ))}{N (1-3 \alpha )^2 (\zeta )^2}\Big{)}\, .
\end{align}
Focusing at the vicinity of the bouncing point $t\rightarrow t_s$,
which correspond to $N\rightarrow 0$, we get,
\begin{align}\label{limitingcasesn0}
& \frac{J_2}{J_1}=3+\frac{3 \zeta }{2 N}\, , \\ \notag &
\frac{J_3}{J_1}=-\mathcal{A}N^{-2+\zeta +\frac{\alpha -2 \alpha
\zeta }{-1+3 \alpha }}\, ,
\end{align}
where the parameter $\mathcal{A}$ is,
\begin{equation}\label{gfredhotlady}
\mathcal{A}=\frac{3^{1+\frac{\alpha }{1-3 \alpha }}
4^{\frac{\alpha }{1-3 \alpha }} \delta (1-3 \alpha )^2 \gamma
\left(\frac{\beta ^4 \gamma }{\delta^2}\right)^{1+\frac{\alpha
}{1-3 \alpha }}}{c_2 \alpha  (-1+2 \alpha ) \beta ^2}\, ,
\end{equation}
and $\mathcal{A}$ is obviously positive. Hence $J_2/J_1>0$ and
$J_3/J_1<0$, therefore the dynamical system of the cosmological
equations is unstable near the bouncing point, as we anticipated.

Now let us consider the stability of the cosmological solution for
the late-time era, focusing at redshifts $z\sim 0$. Since the
late-time evolution results to an EoS for the dark energy
approximately equal to $-1$, the late-time evolution for redshifts
$z\sim 0$, can be approximated by a de Sitter evolution, and it is
realized by the approximate $f(\mathcal{G})$ gravity of the form,
\begin{equation}\label{approximatefg}
f(\mathcal{G})\sim c_1/\mathcal{G}\, .
\end{equation}
Hence assuming that the late-time Hubble rate has the de Sitter
form, $H(t)\sim H_0$, in effect, the function $g(N)$ in this case
has the form $g(N)=H_0^2$. Let us repeat the procedure we followed
for the case of the bounce near the bouncing point, so in this
case, by perturbing the de Sitter vacuum solution $g(N)=H_0^2$
using a linear perturbation of the form (\ref{pert2}), the
variable $J_1$ becomes,
\begin{equation}\label{J1latetime}
J_1=\frac{c_1}{24 H_0^6}\, ,
\end{equation}
while the parameter $J_2$ reads,
\begin{equation}\label{J2latetime}
J_2=\frac{c_1}{8 H_0^6}\, ,
\end{equation}
and finally the parameter $J_3$ reads,
\begin{equation}\label{j3latetime}
J_3=6 \left(-\frac{c_1}{36 H_0^6}+\frac{c_1}{18 H_0^4}+1\right)\,
.
\end{equation}
Now the fraction $J_2/J_1$ is equal to $J_2/J_1=3$, while the
fraction $J_3/J_1$, is equal to,
\begin{equation}\label{fractionj3j1}
\frac{J_3}{J_1}=\frac{144 H_0^6}{c_1}+8 H_0^2-4\, ,
\end{equation}
and since for the late-time analysis we took $H_0\simeq
1.37187\times 10^{-33}$eV and $c_1=1$eV$^6$, this means that
$J_3/J_1$ is obviously negative. This means that the dynamical
system corresponding to the linear cosmological perturbations is
unstable for the late-time solutions. This means that the
late-time dynamical system has an unstable de Sitter attractor
(fixed point), thus in the era that lies beyond $z=0$, the system
will not possibly remain to the de Sitter vacuum, and physically
this could mean that the late-time dark energy EoS will not remain
to the fixed $-1$ value, but might evolve to the phantom or
quintessence regime. In any case, this result is somewhat
interesting, since it deviates from the $\Lambda$CDM description,
in which case the de Sitter state is stable and unchanged, the
cosmological constant is constant.

Before closing, let us briefly discuss the stability of the
cosmological perturbations for the $R+f(\mathcal{G})$ gravity, as
the general relativistic limit is approached. This issue was
covered in detail in Ref. \cite{delaCruzDombriz:2011wn}, so we
shall report their result, which is that the condition
$f''(\mathcal{G})>0$ suffices to ensure the stability of any
solution in the General Relativistic limit. In our case, at the
late-time era, where the $f(\mathcal{G})$ is approximately
$f(\mathcal{G})\sim c_1/\mathcal{G}$, the quantity
$f''(\mathcal{G})$ is equal to $f''(\mathcal(G))=\frac{2
c_1}{\mathcal{G}^3}$, hence the stability is ensured. However, for
the early-time era, in which case the $f(\mathcal{G})$ gravity has
the form $f(\mathcal{G})\sim
c_2\mathcal{G}^{\frac{\alpha}{3\alpha-1}}$, the quantity
$f''(\mathcal{G})$ reads,
\begin{equation}\label{fpp}
f''(\mathcal{G})=-\frac{\alpha  (2 \alpha -1) c_2
\mathcal{G}^{\frac{\alpha }{3 \alpha -1}-2}}{(1-3 \alpha )^2} \, ,
\end{equation}
which is obviously negative for the values of $\alpha$ we chose,
hence this solution is unstable towards to the general
relativistic limit. However, this is not a problem in this case,
because near the bounce era, no consistent general relativistic
limit exists, since general relativity cannot describe the
early-time era, while at the late-time, there can be some overlap
between the general relativistic description and the
$f(\mathcal{G})$ description, because the late-time era is
approximately described by a nearly de Sitter evolution.

\section{Generalized $f(R,\mathcal{G})$ Gravity Late-time Phenomenology: The case of $f(R)+g(\mathcal{G})$ Gravity}

In the previous section we compared the results of the
Gauss-Bonnet late-time phenomenology with the $f(R)$ model of Eq.
(\ref{fR}), and we demonstrated that even though both $f(R)$ and
$R+f(\mathcal{G})$ models are capable of uniting early and
late-time eras, there exist many differences, in particular the
dark energy oscillations spotted on redshifts $z\geq 4$, which are
present in the $f(R)$ gravity model while these are absent in the
Gauss-Bonnet model. It is therefore sensible to try and find out
whether combining the aforementioned $f(R)$ model with an
appropriately chosen $g(\mathcal{G})$ function makes the dark
energy oscillations disappear. Therefore, let us assume that the
$f(R,\mathcal{G})$ model is written as
$f(R,\mathcal{G})=f(R)+g(\mathcal{G})$ with,
\begin{equation}
\centering
\label{fR1}
f(R)=R+\left(\frac{R}{M}\right)^2-\gamma\Lambda\left(\frac{R}{3m_s^2}\right)^\delta\, ,
\end{equation}
and
\begin{equation}
\centering
\label{gG1}
g(\mathcal{G})=\frac{\lambda}{\sqrt{\beta}}\mathcal{G}\tan^{-1}\left(\frac{\mathcal{G}}{\beta}\right)\, ,
\end{equation}
where $M$, $\gamma$ and $\delta$ are the same parameters as in the
previous section, whereas $\lambda$ and $\beta$ are extra
auxiliary parameters, the first being dimensionless while the
latter has mass dimensions $[m]^4$ for consistency and moreover
$\beta$ is assumed to be positive.
\begin{figure}
\centering
\includegraphics[width=20pc]{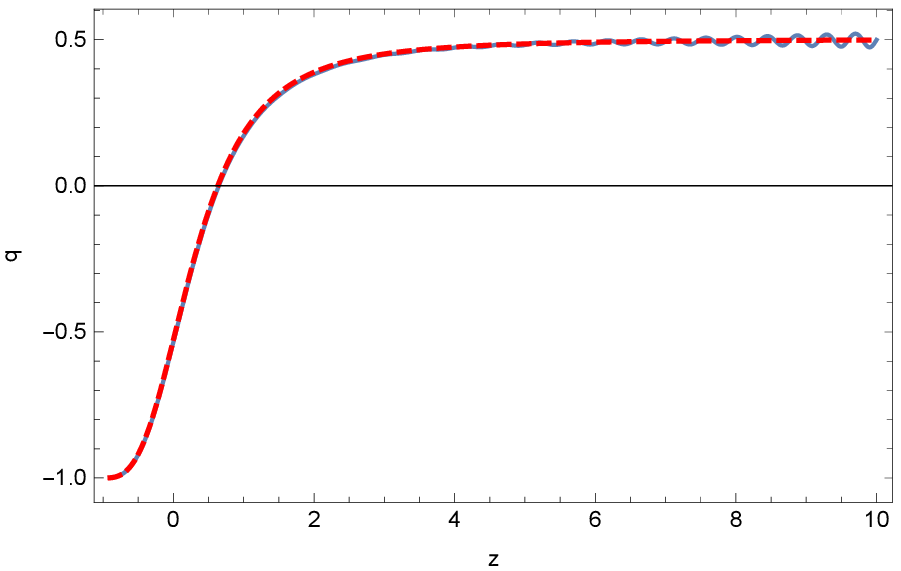}
\includegraphics[width=19.5pc]{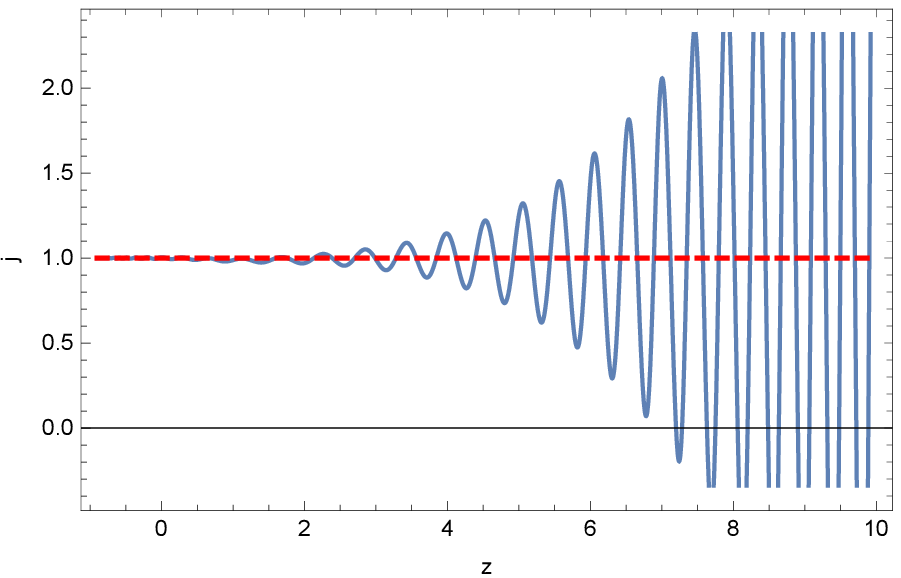}
\includegraphics[width=20pc]{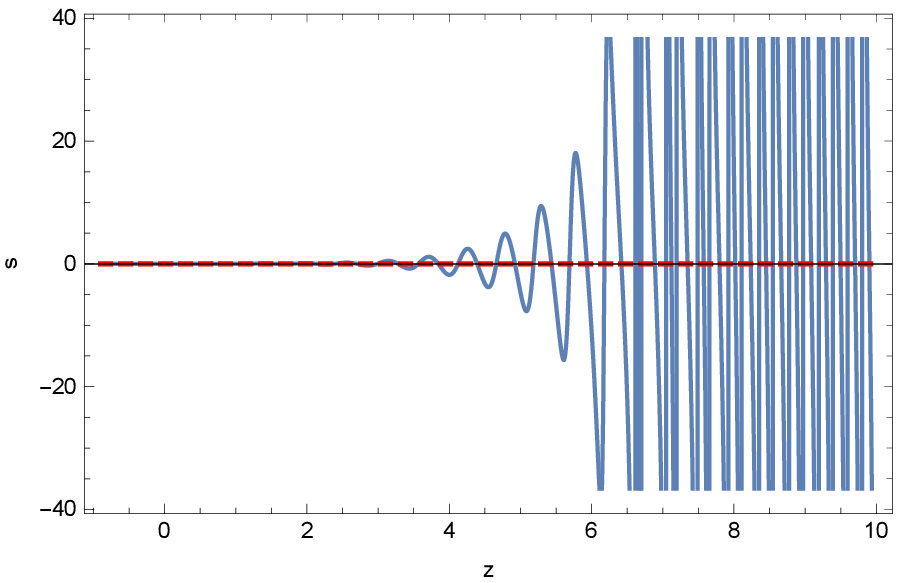}
\includegraphics[width=20pc]{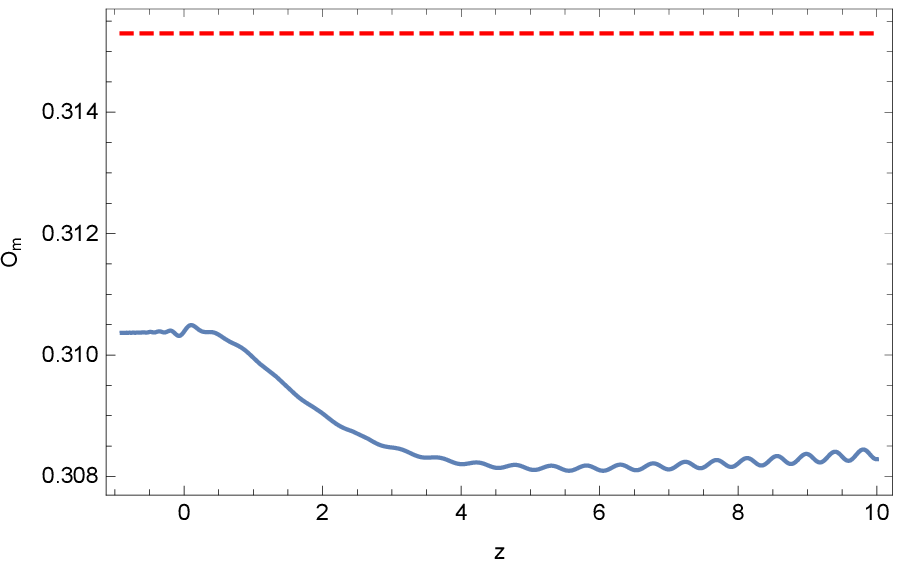}
\caption{Direct comparison of several statefinder quantities
between the $f(R)+g(\mathcal{G})$ model with $\Lambda$CDM. }
\label{plot8}
\end{figure}
The model $g(\mathcal{G})$ seems quite convenient for our approach
since it has a linear part in terms of $\mathcal{G}$ and moreover
the derivative of $\tan^{-1}(\mathcal{G})$ produces a term
$\frac{1}{1+\mathcal{G}^2}$. Since we assume an
$f(R)+g(\mathcal{G})$ gravity, then Eq. (\ref{motion1}) is written
as follows,
\begin{equation}
\centering \label{frgdiff}
3f_RH^2=\kappa^2\rho_{(m)}+\frac{Rf_R-f}{2}-3H\dot
f_{R}+\frac{\mathcal{G}g_\mathcal{G}-g}{2}-12\dot
g_{\mathcal{G}}H^3\, .
\end{equation}
This is the Friedmann equation, and this particular differential
equation shall be solved numerically in the same redshift interval
$z=[-0.9,10]$, by expressing the above differential equation in
terms of the statefinder quantity $y_H$. In the case at hand, we
have,
\begin{equation}
\centering
\label{FR}
f_R=1+\frac{2R}{M^2}-\frac{\gamma\delta\Lambda}{(3m_s^2)^\delta}R^{\delta-1}\, ,
\end{equation}
\begin{equation}
\centering
\label{FRR}
\dot f_R=\dot R\left(\frac{2}{M^2}-\frac{\gamma\delta(1-\delta)\Lambda}{(3m_s^2)^\delta}R^{\delta-2}\right)\, ,
\end{equation}
\begin{equation}
\centering
\label{FG}
g_\mathcal{G}=\frac{\lambda}{\sqrt{\beta}}\left(\tan^{-1}\left(\frac{\mathcal{G}}{\beta}\right)+\frac{\beta\mathcal{G}}{\beta^2+\mathcal{G}^2}\right)\, ,
\end{equation}
\begin{equation}
\centering \label{gGG} \dot g_\mathcal{G}=2\sqrt{\beta} \lambda
\mathcal{\dot
G}\frac{\mathcal{G}}{\beta^2+\mathcal{G}^2}\left(1-\frac{\mathcal{G}}{\beta^2+\mathcal{G}^2}\right)\,
.
\end{equation}
Suppose now that the values for $M$, $m_s$, $\gamma$, $\delta$,
$\Lambda$ remain the same as in the previous section, hence the
reason they were not relabelled, and in addition $\lambda=1$,
$\beta=1$. Then by using exactly the same initial conditions, and
by solving numerically the differential equation (\ref{frgdiff}),
we present the results of our analysis in Figs. \ref{plot6},
\ref{plot7} and \ref{plot8}.

Particularly, in Fig. \ref{plot6} we plot the deceleration
parameter and and statefinder $y_H$ as functions of the redshift,
and in Fig. \ref{plot7} we plot the dark energy EoS parameter and
the dark energy density parameter as functions of the redshift.
Finally, in Fig. \ref{plot8} we plot several statefinders for the
combined $f(R)+g(\mathcal{G})$ model (blue curves) and we compare
these to the $\Lambda$CDM model results (red curves). Also in
Table \ref{table2}, we compare the values of several quantities of
cosmological interest at present time for the combined
$f(R)+g(\mathcal{G})$ model, and compare these to the pure $f(R)$
gravity model. An overall apparent result derived from the
analysis, is that the presence of the $f(R)$ gravity utterly
affects the late-time phenomenology, since it brings along the
dark energy oscillations in several statefinder quantities and
cosmological quantities. This result seems to be
model-independent, thus in conclusion, the $f(R)$ gravity dark
energy oscillations cannot be remedied by adding a Gauss-Bonnet
term in the Lagrangian.
\begin{table}[h!]
\label{table2} \caption{ Comparison between the values of pure
$f(R)$ model and of the the $f(R)+g(\mathcal{G})$ case. Even
though all the statefinders seem to have the same behavior in the
redshift interval $z=[-0.9,10]$, the current values are indeed
different, however to a small extent.}
\begin{center}
\begin{tabular}{|r|r|r|}
\hline
\textbf{Parameter}&\textbf{$f(R)$ Value}&\textbf{$f(R)+g(\mathcal{G})$ Value}\\ \hline
q(z=0)&-0.520954&-0.53442\\ \hline
j(z=0)&1.00319&1.00478\\ \hline
s(z=0)&-0.00104169&-0.00154146\\ \hline
$Om(z=0)$&0.319364&0.310387\\ \hline
$\Omega_{DE}(0)$&0.683948&0.691643\\ \hline
$\omega_{DE}(0)$&-0.995205&-0.995673\\ \hline
\end{tabular}
\end{center}
\end{table}

\section{Conclusions}

In this paper we quantitatively addressed the late-time
phenomenology of Gauss-Bonnet theories, and we investigated how a
singular bouncing cosmology occurring primordially and a viable
dark energy era can be realized by Gauss-Bonnet gravity. This
unification scheme was possible to be realized by using a
Gauss-Bonnet gravity of the form $R+f(\mathcal{G})$, which is free
from primordial superluminal modes. The Type IV singular bounce at
early times, realized by an appropriate Gauss-Bonnet gravity,
generates a nearly scale invariant power spectrum of the
primordial scalar curvature perturbations. In addition, the
late-time driving part of the $R+f(\mathcal{G})$ gravity generates
a viable dark energy era, which mimics the $\Lambda$CDM model for
some statefinder quantities, and is compatible at present time
with the latest Planck data on cosmological parameters. This
specific model, has another appealing attribute, the fact that the
dark energy era is free from large redshift ($z \sim 4$) dark
energy oscillations, known to occur in $f(R)$ theories. For
demonstrative reasons, we compared the $R+f(\mathcal{G})$ with the
$\Lambda$CDM model and an appropriate $f(R)$ gravity model which
is known to generate a viable dark energy, and we provided
qualitative evidence for the presence of dark energy oscillations
only in the $f(R)$ gravity case. In order to further analyze the
impact of a non-trivial $f(R)$ gravity term in the dark energy
oscillations, we also studied the late-time phenomenology of an
$f(R)+g(\mathcal{G})$, and as we showed, the dark energy
oscillations are also present in this case. Thus the $f(R)$
gravity part in the gravitational action seems to always lead to
dark energy oscillations at large redshifts.


\begin{thebibliography}{99}










\bibitem{Bertone:2004pz}
  G.~Bertone, D.~Hooper and J.~Silk,
  Phys.\ Rept.\  {\bf 405} (2005) 279
  doi:10.1016/j.physrep.2004.08.031
  [hep-ph/0404175].


\bibitem{Bergstrom:2000pn}
  L.~Bergstrom,
  Rept.\ Prog.\ Phys.\  {\bf 63} (2000) 793
  doi:10.1088/0034-4885/63/5/2r3
  [hep-ph/0002126].




\bibitem{Mambrini:2015sia}
  Y.~Mambrini, S.~Profumo and F.~S.~Queiroz,
  Phys.\ Lett.\ B {\bf 760} (2016) 807
  [arXiv:1508.06635 [hep-ph]].

\bibitem{Profumo:2013yn}
  S.~Profumo,
  arXiv:1301.0952 [hep-ph].




\bibitem{Hooper:2007qk}
  D.~Hooper and S.~Profumo,
  Phys.\ Rept.\  {\bf 453} (2007) 29
  [hep-ph/0701197].




\bibitem{Oikonomou:2006mh}
V.~K.~Oikonomou, J.~D.~Vergados and C.~C.~Moustakidis,
Nucl.\ Phys.\ B {\bf 773} (2007) 19
[hep-ph/0612293].



\bibitem{Capozziello:2012ie}
S.~Capozziello and M.~De Laurentis,
Annalen Phys. \textbf{524} (2012), 545-578
doi:10.1002/andp.201200109









\bibitem{Riess:1998cb}
  A.~G.~Riess {\it et al.} [Supernova Search Team],
  Astron.\ J.\  {\bf 116} (1998) 1009
  doi:10.1086/300499
  [astro-ph/9805201].



\bibitem{Bamba:2012cp}
  K.~Bamba, S.~Capozziello, S.~Nojiri and S.~D.~Odintsov,
  Astrophys.\ Space Sci.\  {\bf 342} (2012) 155
  doi:10.1007/s10509-012-1181-8
  [arXiv:1205.3421 [gr-qc]].


\bibitem{Peebles:2002gy}
  P.~J.~E.~Peebles and B.~Ratra,
  Rev.\ Mod.\ Phys.\  {\bf 75} (2003) 559
  doi:10.1103/RevModPhys.75.559
  [astro-ph/0207347].


\bibitem{Li:2011sd}
  M.~Li, X.~D.~Li, S.~Wang and Y.~Wang,
  Commun.\ Theor.\ Phys.\  {\bf 56} (2011) 525
  doi:10.1088/0253-6102/56/3/24
  [arXiv:1103.5870 [astro-ph.CO]].


\bibitem{Bamba:2010wb}
  K.~Bamba, C.~Q.~Geng, C.~C.~Lee and L.~W.~Luo,
  JCAP {\bf 1101} (2011) 021
  doi:10.1088/1475-7516/2011/01/021
  [arXiv:1011.0508 [astro-ph.CO]].

\bibitem{Frieman:2008sn}
  J.~Frieman, M.~Turner and D.~Huterer,
  Ann.\ Rev.\ Astron.\ Astrophys.\  {\bf 46} (2008) 385
  doi:10.1146/annurev.astro.46.060407.145243
  [arXiv:0803.0982 [astro-ph]].


\bibitem{Boehmer:2008av}
  C.~G.~Boehmer, G.~Caldera-Cabral, R.~Lazkoz and R.~Maartens,
  Phys.\ Rev.\ D {\bf 78} (2008) 023505
  doi:10.1103/PhysRevD.78.023505
  [arXiv:0801.1565 [gr-qc]].





\bibitem{Nojiri:2006gh}
  S.~Nojiri and S.~D.~Odintsov,
  Phys.\ Rev.\ D {\bf 74} (2006) 086005
  doi:10.1103/PhysRevD.74.086005
  [hep-th/0608008].

\bibitem{Elizalde:2004mq}
  E.~Elizalde, S.~Nojiri and S.~D.~Odintsov,
  Phys.\ Rev.\ D {\bf 70} (2004) 043539
  doi:10.1103/PhysRevD.70.043539
  [hep-th/0405034].



\bibitem{Makarenko:2018blx}
  A.~N.~Makarenko and A.~N.~Myagky,
  Int.\ J.\ Geom.\ Meth.\ Mod.\ Phys.\  {\bf 15} (2018) no.06,  1850096.
  doi:10.1142/S0219887818500962










\bibitem{Capozziello:2003gx}
  S.~Capozziello, V.~F.~Cardone, S.~Carloni and A.~Troisi,
  Int.\ J.\ Mod.\ Phys.\ D {\bf 12} (2003) 1969
  doi:10.1142/S0218271803004407
  [astro-ph/0307018].


\bibitem{Kamenshchik:2001cp}
  A.~Y.~Kamenshchik, U.~Moschella and V.~Pasquier,
  Phys.\ Lett.\ B {\bf 511} (2001) 265
  doi:10.1016/S0370-2693(01)00571-8
  [gr-qc/0103004].


\bibitem{Carroll:1998zi}
  S.~M.~Carroll,
  Phys.\ Rev.\ Lett.\  {\bf 81} (1998) 3067
  doi:10.1103/PhysRevLett.81.3067
  [astro-ph/9806099].







\bibitem{Capozziello:2002rd}
  S.~Capozziello,
  Int.\ J.\ Mod.\ Phys.\ D {\bf 11} (2002) 483
  doi:10.1142/S0218271802002025
  [gr-qc/0201033].






\bibitem{Capozziello:2005ra}
  S.~Capozziello, V.~F.~Cardone, E.~Piedipalumbo and C.~Rubano,
  Class.\ Quant.\ Grav.\  {\bf 23} (2006) 1205
  doi:10.1088/0264-9381/23/4/009
  [astro-ph/0507438].








\bibitem{reviews1}
 S.~Nojiri, S.~D.~Odintsov and V.~K.~Oikonomou,
  Phys.\ Rept.\  {\bf 692} (2017) 1
  doi:10.1016/j.physrep.2017.06.001
  [arXiv:1705.11098 [gr-qc]].

\bibitem{reviews2}

S. Nojiri, S.D. Odintsov,
   Phys.\ Rept.\  {\bf 505}, 59 (2011);


  \bibitem{reviews3}
S. Nojiri, S.D. Odintsov,
  eConf {\bf C0602061}, 06 (2006)
  [Int.\ J.\ Geom.\ Meth.\ Mod.\ Phys.\  {\bf 4}, 115 (2007)].


   \bibitem{reviews4}
 S. Capozziello, M. De Laurentis,
   Phys.\ Rept.\  {\bf 509}, 167 (2011);\\
 V.~Faraoni and S.~Capozziello,
  Fundam.\ Theor.\ Phys.\  {\bf 170} (2010).
  doi:10.1007/978-94-007-0165-6



\bibitem{reviews5}

A.~de la Cruz-Dombriz and D.~Saez-Gomez,
  Entropy {\bf 14} (2012) 1717
  doi:10.3390/e14091717
  [arXiv:1207.2663 [gr-qc]].

\bibitem{reviews6}

G.~J.~Olmo,
  Int.\ J.\ Mod.\ Phys.\ D {\bf 20} (2011) 413
  doi:10.1142/S0218271811018925
  [arXiv:1101.3864 [gr-qc]].



\bibitem{Guth:1980zm}
  A.~H.~Guth,
  Phys.\ Rev.\ D {\bf 23} (1981) 347.
  doi:10.1103/PhysRevD.23.347


\bibitem{Linde:1993cn}
  A.~D.~Linde,
  Phys.\ Rev.\ D {\bf 49} (1994) 748
  doi:10.1103/PhysRevD.49.748
  [astro-ph/9307002].


\bibitem{Linde:1983gd}
  A.~D.~Linde,
  Phys.\ Lett.\  {\bf 129B} (1983) 177.
  doi:10.1016/0370-2693(83)90837-7








\bibitem{Brandenberger:2016vhg}
  R.~Brandenberger and P.~Peter,
  Found.\ Phys.\  {\bf 47} (2017) no.6,  797
  doi:10.1007/s10701-016-0057-0
  [arXiv:1603.05834 [hep-th]].;\\
\bibitem{deHaro:2015wda}
  J.~de Haro and Y.~F.~Cai,
  Gen.\ Rel.\ Grav.\  {\bf 47} (2015) no.8,  95
  doi:10.1007/s10714-015-1936-y
  [arXiv:1502.03230 [gr-qc]].;\\
\bibitem{Cai:2014bea}
  Y.~F.~Cai,
  Sci.\ China Phys.\ Mech.\ Astron.\  {\bf 57} (2014) 1414
  doi:10.1007/s11433-014-5512-3
  [arXiv:1405.1369 [hep-th]].

\bibitem{Avelino:2012ue}

P.~P.~Avelino and R.~Z.~Ferreira,
Phys. Rev. D \textbf{86} (2012), 041501
doi:10.1103/PhysRevD.86.041501 [arXiv:1205.6676 [astro-ph.CO]].




\bibitem{Koehn:2013upa}

M.~Koehn, J.~L.~Lehners and B.~A.~Ovrut,
Phys. Rev. D \textbf{90} (2014) no.2, 025005
doi:10.1103/PhysRevD.90.025005 [arXiv:1310.7577 [hep-th]].




\bibitem{Cai:2013kja}
Y.~F.~Cai, E.~McDonough, F.~Duplessis and R.~H.~Brandenberger,
JCAP \textbf{10} (2013), 024 doi:10.1088/1475-7516/2013/10/024
[arXiv:1305.5259 [hep-th]].

\bibitem{Brandenberger:2012zb}
R.~H.~Brandenberger,
[arXiv:1206.4196 [astro-ph.CO]].


\bibitem{Cai:2011zx}
Y.~F.~Cai, R.~Brandenberger and X.~Zhang,
JCAP \textbf{03} (2011), 003 doi:10.1088/1475-7516/2011/03/003
[arXiv:1101.0822 [hep-th]].


\bibitem{Allen:2004vz}
L.~E.~Allen and D.~Wands,
Phys. Rev. D \textbf{70} (2004), 063515
doi:10.1103/PhysRevD.70.063515 [arXiv:astro-ph/0404441
[astro-ph]].





\bibitem{Nojiri:2003ft}
  S.~Nojiri and S.~D.~Odintsov,
  Phys.\ Rev.\ D {\bf 68} (2003) 123512
  doi:10.1103/PhysRevD.68.123512
  [hep-th/0307288].



\bibitem{Odintsov:2020nwm}
S.~D.~Odintsov and V.~K.~Oikonomou,
Phys. Rev. D \textbf{101} (2020) no.4, 044009
doi:10.1103/PhysRevD.101.044009 [arXiv:2001.06830 [gr-qc]].






\bibitem{Caldwell:2003vq}
  R.~R.~Caldwell, M.~Kamionkowski and N.~N.~Weinberg,
  Phys.\ Rev.\ Lett.\  {\bf 91} (2003) 071301
  doi:10.1103/PhysRevLett.91.071301
  [astro-ph/0302506].




\bibitem{Li:2007jm}
  B.~Li, J.~D.~Barrow and D.~F.~Mota,
``The Cosmology of Modified Gauss-Bonnet Gravity,''
  Phys.\ Rev.\ D {\bf 76} (2007) 044027
  doi:10.1103/PhysRevD.76.044027
  [arXiv:0705.3795 [gr-qc]].




\bibitem{Nojiri:2005jg}
 S.~Nojiri and S.~D.~Odintsov,
``Modified Gauss-Bonnet theory as gravitational alternative for
dark energy,''
 Phys.\ Lett.\ B {\bf 631} (2005) 1
 [hep-th/0508049].




\bibitem{Elizalde:2020zcb}
  E.~Elizalde, S.~D.~Odintsov, V.~K.~Oikonomou and T.~Paul,
  Nucl.\ Phys.\ B {\bf 954} (2020) 114984
  doi:10.1016/j.nuclphysb.2020.114984
  [arXiv:2003.04264 [gr-qc]].

\bibitem{Cognola:2006eg}
 G.~Cognola, E.~Elizalde, S.~Nojiri, S.~D.~Odintsov and S.~Zerbini,
``Dark energy in modified Gauss-Bonnet gravity: Late-time
acceleration and the hierarchy problem,''
 Phys.\ Rev.\ D {\bf 73} (2006) 084007
 [hep-th/0601008].

\bibitem{Elizalde:2010jx}
 E.~Elizalde, R.~Myrzakulov, V.~V.~Obukhov and D.~Saez-Gomez,
``LambdaCDM epoch reconstruction from $F(R,\mathcal{G})$ and
modified Gauss-Bonnet gravities,''
 Class.\ Quant.\ Grav.\ {\bf 27} (2010) 095007
 [arXiv:1001.3636 [gr-qc]].





\bibitem{Izumi:2014loa}
 K.~Izumi,
``Causal Structures in Gauss-Bonnet gravity,''
 Phys.\ Rev.\ D {\bf 90} (2014) no.4, 044037
 [arXiv:1406.0677 [gr-qc]].


\bibitem{Oikonomou:2016rrv}
  V.~K.~Oikonomou,
``Gauss-Bonnet Cosmology Unifying Late and Early-time Acceleration
Eras with Intermediate Eras,''
  Astrophys.\ Space Sci.\  {\bf 361} (2016) no.7,  211
  doi:10.1007/s10509-016-2800-6
  [arXiv:1606.02164 [gr-qc]].

\bibitem{Kleidis:2017ftt}
  K.~Kleidis and V.~K.~Oikonomou,
  Int.\ J.\ Geom.\ Meth.\ Mod.\ Phys.\  {\bf 15} (2017) no.04,  1850064
  doi:10.1142/S0219887818500640
  [arXiv:1711.09270 [gr-qc]].

\bibitem{Oikonomou:2015qha}
  V.~K.~Oikonomou,
``Singular Bouncing Cosmology from Gauss-Bonnet Modified
Gravity,''
  Phys.\ Rev.\ D {\bf 92} (2015) no.12,  124027
  doi:10.1103/PhysRevD.92.124027
  [arXiv:1509.05827 [gr-qc]].



\bibitem{Escofet:2015gpa}
  A.~Escofet and E.~Elizalde,
``Gauss-Bonnet modified gravity models with bouncing behavior,''
  Mod.\ Phys.\ Lett.\ A {\bf 31} (2016) no.17,  1650108
  doi:10.1142/S021773231650108X
  [arXiv:1510.05848 [gr-qc]].





\bibitem{new2}

 K.~Bamba, A.~N.~Makarenko, A.~N.~Myagky and S.~D.~Odintsov,
``Bouncing cosmology in modified Gauss-Bonnet gravity,''
  Phys.\ Lett.\ B {\bf 732} (2014) 349
  doi:10.1016/j.physletb.2014.04.004
  [arXiv:1403.3242 [hep-th]].


\bibitem{Makarenko:2016jsy}
  A.~N.~Makarenko,
``The role of Lagrange multiplier in Gauss-Bonnet dark energy,''
  Int.\ J.\ Geom.\ Meth.\ Mod.\ Phys.\  {\bf 13} (2016) no.05,  1630006.
  doi:10.1142/S0219887816300063




\bibitem{Navo:2020eqt}
G.~Navo and E.~Elizalde,
doi:10.1142/S0219887820501625 [arXiv:2007.11507 [gr-qc]].


\bibitem{Bajardi:2020osh}
F.~Bajardi and S.~Capozziello,
Eur. Phys. J. C \textbf{80} (2020) no.8, 704
doi:10.1140/epjc/s10052-020-8258-2 [arXiv:2005.08313 [gr-qc]].


\bibitem{Capozziello:2019wfi}
S.~Capozziello, C.~A.~Mantica and L.~G.~Molinari,
Int. J. Geom. Meth. Mod. Phys. \textbf{16} (2019) no.09, 1950133
doi:10.1142/S0219887819501330 [arXiv:1906.05693 [gr-qc]].


\bibitem{Benetti:2018zhv}
M.~Benetti, S.~Santos da Costa, S.~Capozziello, J.~S.~Alcaniz and
M.~De Laurentis,
Int. J. Mod. Phys. D \textbf{27} (2018) no.08, 1850084
doi:10.1142/S0218271818500840 [arXiv:1803.00895 [gr-qc]].







\bibitem{Clifton:2006kc}
  T.~Clifton and J.~D.~Barrow,
  Class.\ Quant.\ Grav.\  {\bf 23} (2006) 2951
  doi:10.1088/0264-9381/23/9/011
  [gr-qc/0601118].



\bibitem{Bogdanos:2009tn}
  C.~Bogdanos, S.~Capozziello, M.~De Laurentis and S.~Nesseris,
  Astropart.\ Phys.\  {\bf 34} (2010) 236
  doi:10.1016/j.astropartphys.2010.08.001
  [arXiv:0911.3094 [gr-qc]].


\bibitem{Capozziello:2004us}
  S.~Capozziello, V.~F.~Cardone, S.~Carloni and A.~Troisi,
  Phys.\ Lett.\ A {\bf 326} (2004) 292
  doi:10.1016/j.physleta.2004.04.081
  [gr-qc/0404114].







\bibitem{Bamba:2009uf}
  K.~Bamba, S.~D.~Odintsov, L.~Sebastiani and S.~Zerbini,
  Eur.\ Phys.\ J.\ C {\bf 67} (2010) 295
  doi:10.1140/epjc/s10052-010-1292-8
  [arXiv:0911.4390 [hep-th]].




\bibitem{DeLaurentis:2015fea}
  M.~De Laurentis, M.~Paolella and S.~Capozziello,
  Phys.\ Rev.\ D {\bf 91} (2015) no.8,  083531
  doi:10.1103/PhysRevD.91.083531
  [arXiv:1503.04659 [gr-qc]].


\bibitem{delaCruzDombriz:2011wn}
  A.~de la Cruz-Dombriz and D.~Saez-Gomez,
  Class.\ Quant.\ Grav.\  {\bf 29} (2012) 245014
  doi:10.1088/0264-9381/29/24/245014
  [arXiv:1112.4481 [gr-qc]].

\bibitem{Aghanim:2018eyx}
N.~Aghanim \textit{et al.} [Planck],
[arXiv:1807.06209 [astro-ph.CO]].




\bibitem{Akrami:2018odb}
  Y.~Akrami {\it et al.} [Planck Collaboration],
  arXiv:1807.06211 [astro-ph.CO].



\bibitem{Bamba:2014mya}
K.~Bamba, A.~N.~Makarenko, A.~N.~Myagky and S.~D.~Odintsov,
Phys. Lett. B \textbf{732} (2014), 349-355



\bibitem{delaCruzDombriz:2011wn}
A.~de la Cruz-Dombriz and D.~Saez-Gomez,
Class. Quant. Grav. \textbf{29} (2012), 245014




\end{thebibliography}
\end{document}